%% file: paper_movContLine_PF_MD_VOF.tex
\documentclass[epjST]{svjour}
\usepackage{amsmath}
\usepackage{graphicx}

\usepackage{tikz}
\usepackage{comment}
\usepackage[caption = false]{subfig}
\usepackage{longtable}
\usepackage{multirow}

\newcommand{\rev}[1]{{\color{black}#1}}

\newcommand{\SiO}{SiO${}_2$}

\newcommand{\pd}{\partial}

\newcommand{\staticContactAngle}{97.5}

\renewcommand\Re{{\rm Re}\,}

\newcommand\Ca{{\rm Ca}\,}

\newcommand\Pe{{\rm Pe}\,}

\newcommand\Cn{C_n}
\input{macros.tex}

\begin{document}
\title{Steady moving contact line of water over a no-slip substrate}
\subtitle{Challenges in benchmarking phase-field and volume-of-fluid methods against molecular
dynamics simulations}
\author{U\v{g}is L\={a}cis\inst{1}\fnmsep\thanks{\email{ugis@mech.kth.se}}
\and Petter Johansson\inst{2}
\and Tomas Fullana\inst{3}
\and Berk Hess\inst{2}
\and Gustav Amberg\inst{1,4}
\and Shervin Bagheri\inst{1}
\and Stephan\'{e} Zaleski\inst{1,3}}
\institute{FLOW centre, Department of Engineering Mechanics KTH, SE-100 44 Stockholm, Sweden
\and Swedish e-Science Research Centre, Science for Life Laboratory, Department of Applied Physics KTH, SE-100 44 Stockholm, Sweden
\and Sorbonne Universit\'{e} and CNRS, France
\and S\"{o}dertorn University, Stockholm, Sweden}
\abstract{
The movement of the triple contact line
plays a crucial role in many applications
such as ink-jet printing, liquid coating and drainage (imbibition) in porous media.
To design accurate computational tools for these applications, predictive models of the moving contact line are needed. However, the basic mechanisms
responsible for movement of the triple contact line are not well understood
but still debated. We investigate the movement of the contact line between
water, vapour and a silica-like solid surface under steady conditions in
low capillary number regime. We use molecular dynamics (MD) with an atomistic water
model to simulate a nanoscopic drop between two moving plates. We include hydrogen bonding between the water molecules and
the solid substrate, which leads to a sub-molecular slip length.
We benchmark two continuum methods, the Cahn--Hilliard phase-field (PF)
model and a volume-of-fluid (VOF) model, against MD results. We show that both continuum models
reproduce the statistical measures obtained from MD reasonably well, with a
trade-off in accuracy. We demonstrate the importance of the
phase-field mobility parameter and the local slip length
in accurately modelling the moving contact line.
} 
\maketitle
\section{Introduction}
\label{intro}

The motion of a two-fluid interface contacting a flat solid surface poses a
particularly difficult problem of continuum fluid mechanics. If the
traditional point of view of a no-slip wall -- a sharp transition between the phases
and constant surface tension -- is to be believed, then a contradiction
ensues since at the triple point or contact line the velocity is both
zero and non zero \cite{huh1971hydrodynamic}. Attempts to solve this paradox and make
progress on the issue abound \cite{bonn:was,snoeijer2013moving,Sui2014}. One
of the most popular is the assumption of Navier slip \cite{navier1823memoire},
but in general all solutions to the paradox amount to the introduction
of a small length scale $l_\mu$ below which the continuum
model ceases to be valid as discussed by Voinov \cite{voinov1976}. Cox \cite{cox1986dynamics} extended
Voinov's theory to arbitrary viscosity ratios and contact angles.
Since then, many theoretical endeavours has been directed towards solving
this problem \cite{Shikhmurzaev1994,shikhmurzaev1997moving,kalliadasis2000steady,eggers2004hydrodynamic,flitton2004surface,snoeijer2006free,pismen2008solvability,snoeijer2010asymptotic,chan2012theory,nold2018vicinity} and the effort continues.
%
%
Nevertheless, the microscopic scale physics remain
somewhat mysterious as described for example in the review by
T. D. Blake \cite{Blake:2015jn}. Indeed,
and beyond the paradox described above, there are many uncertain features
of the nanoscopic flow and interface shape:
there is uncertainty about the value of the contact angle at the smallest scale, and
about the nature of the deviation from equilibrium,
the effect of molecular forces, the presence of evaporation, etc.
Despite recent advancements \cite{deng2016nanoscale},
experiments have difficulty providing interface shape and velocity field
data for moving contact lines below the micron scale.
From the applied mathematical point of view,
Navier slip regularisation leads to an
approximation in which the curvature still diverges logarithmically
at the contact line and to a contradiction if the velocity field is continuous
\cite{Fricke:2018cs,fricke2019kinematic}. Moreover certain fluid
and surface combinations have very small
slip lengths, below the nanometer scale \cite{johansson2015water,johansson2018molecular}.
In such systems, if one considers the problem
at smaller and smaller scales, other molecular effects will become relevant
before the slippage effects.

Thus a full experimental characterisation at the nanometer scale is
difficult and it very well may be that, at least for some
time, only  molecular dynamics (MD) ``numerical
experiments'' will provide the insight necessary to understand which
regularisation is adequate, and particularly so when the slip length
is small and other effects than slip are required for a well-posed problem.
However the largest possible MD
simulations in three dimensions are limited in physical size to only tens of
nanometers in each direction. This would make it impossible 
to perform asymptotic matching
with numerical solutions of the
Navier--Stokes equations. Indeed, in order to perform the matching over all
scales necessary, the Navier--Stokes equations would have to be
solved down to the nanometer scale.
Moreover, for a 1 mm droplet, refining
the calculation to the nanometer scale would require a range of
scales of $10^6$, which implies the impossible-to-attain number of
$10^{18}$ grid points. Even restricting the computations to two dimensions of space,
$10^{12}$ grid points sit on the borderline of currently feasible
computations (and not all problems warrant the use of a
supercomputer).
%
This creates the necessity either for investigations limited to much smaller
scales, \rev{technically adapted approaches to to enable 
the hybrid method for matching MD and Navier-Stokes solvers \cite{hadjiconstantinou1999hybrid,zhang2017multiscale,smith2018moving,borg2018multiscale},}
or for an intermediate-scale model, between the molecular scales and
the scale of the sharp-interface model.
Such an intermediate scale model
may be provided by the diffuse interface approach, in which the hypothesis
of a sharp transition between phases is replaced by that of a smooth
transition over a small length scale $\epsilon$.
In the words of L. Pismen \cite{pismen2011discussion}
``{\em Multiscale methods employing different techniques at disparate length and time scales is the only feasible way to overcome the scale gap in practical computations}''.
Simultaneous usage of MD, a diffuse interface model and a sharp interface
model would be an implementation of this program,
with the diffuse interface approach playing a key role at the intermediate scale.
In such a framework, MD would help design and define parameters for
the diffuse interface model, which in turn would perform a similar service for
larger scale sharp interface models.

Beyond offering an intermediate scale, diffuse interface models open
the possibility of regularising the contact line problem in an efficient and physically
meaningful way.
%
%
Indeed, together with MD, they have led to a
generalisation of the Navier slip boundary condition (so called GNBC) that takes into
account the ``uncompensated Young's stress''\footnote{The inception
of the GNBC is sometimes attributed to T. D. Blake \cite{blake1993dynangleChapter},
who discuss the uncompensated Young's stress and slippage velocity in context
of adsorption and desorption modelling.} and makes the contact line
motion problem well posed \cite{qian2003molecular,QIAN:2006hl}.
On the other hand, diffuse interface models
\cite{carlson2011dissipation,jacqmin2000contact} introduce at least
two additional length scales,
the interface thickness $\eps=\beta/\alpha$ and the diffusion
length $l_d = \sqrt{\mu_l M}$ (the reader is referred to
section~\ref{sec:pf}
and appendix~\ref{app:pf} for a presentation of
the diffusion-based phase-field (PF) model parameters). The number of length
scales is different, for example, for van der Waals
model \cite{laurila2012thermohydrodynamics} or Cahn--Allen
model \cite{eggleston2001phase}.
Since the newly introduced length scales come in
addition of the slip
length $l_s$ there are potentially three length scales that can play an important
role in the physics of the contact line.
In addition a diffuse interface model has several other parameters specifically
related to the contact line: the so called contact line friction
\cite{carlson2009modeling,carlson2011dissipation},
the surface energies (that lead to the equilibrium contact angle through Young's relation)
and their spatial distribution. This makes the selection of
the parameters rather difficult: there are three length scales and
two or more other parameters to adjust or select. In a way, this is the price to pay to
have a more realistic continuum description of the nanoscale.

\begin{figure}
    \centering
    \includegraphics[width=1.0\linewidth]{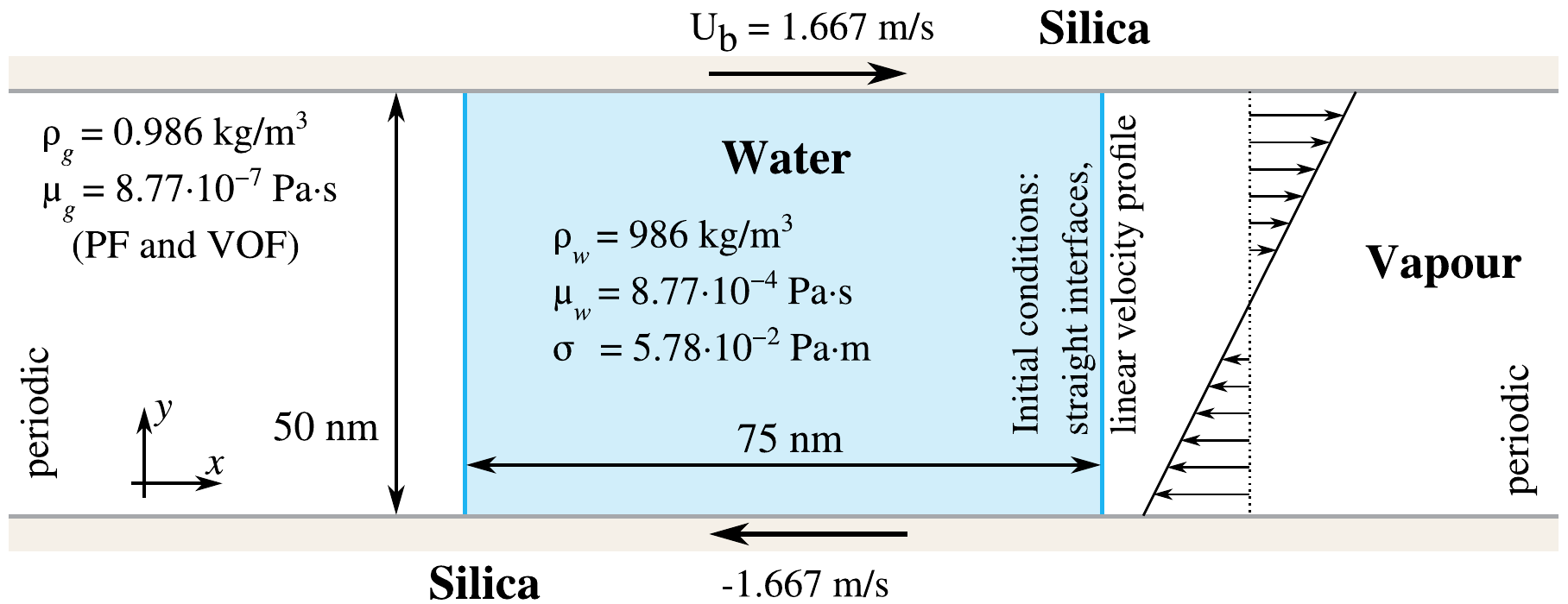}
    \caption{Illustration of the two-phase forced wetting
    configuration. Water parameters are
    determined from the molecular dynamics water model, 
    at temperature $T = 300$ K. The vapour parameters for phase
    field and volume-of-fluid models are $1000$
    times smaller. The chosen parameters lead to capillary number
    $\Ca{} = 2 \mu_w U_b / \sigma = 0.05$.}
    \label{fig:set-up}
\end{figure}

This motivates our choice of a physical situation where two
simplifications occur: small slip
and negligible evaporation.
Indeed, for water on silica far from the critical point, slip is relatively small and
evaporation is moderate. This situation may be reproduced using
MD based on the SPC/E water model, which allows us to
capture the strong hydrogen bonds
between the water molecules and silica
molecules on the surface \cite{johansson2015water}.
Many possible application targeted configurations can be investigated,
such as capillary driven flows \cite{villanueva2006some} or
forced wetting flows \cite{Blake:2015jn}. For the purpose of this work,
we use a two-dimensional water drop enclosed by two moving walls,
as shown in Fig.~\ref{fig:set-up} at sufficiently small capillary number
$\Ca{} = 2 \mu_w U_b / \sigma = 0.05$ for existence of steady state
configuration \cite{jacqmin2004onset,sbragaglia2008wetting}.
Here, $U_b$ is wall velocity, $\mu_w$ is water viscosity
and $\sigma$ is surface tension.
The choice of water in contact with a silica-line substrate
makes the work reported here markedly different from
previous MD investigations
of the contact line dynamics in the same geometrical
configuration \cite{qian2003molecular,ren2007boundary,Blake:2015jn}.
The chose geometrical configuration
is also known for its simplicity: it involves a symmetrical setting with identical
solid walls providing us with identical two phase interfaces.
Using this configuration we report on a first attempt
to match MD, PF, and
volume-of-fluid (VOF) simulations.
By matching we mean finding the PF or VOF parameters
that best reproduces the steady
shape of the interface and the steady velocity field from the
MD.
This procedure is, as far as we know, original. Several other
studies of the correspondence between the PF parameters \cite{barclay2019cahn}
\rev{or the VOF parameters \cite{mohand2019use}}
and MD were performed, but not
the direct attack on the contact line dynamics problem
on a no-slip substrate as we suggest here.

This paper is organised as follows. In section~\ref{sec:molecular-dynamics}
we describe the MD simulations and the measurements, later used for
the benchmarking of PF and VOF. Then, in section~\ref{sec:pf} we describe
the PF model we use, the matching procedure to reproduce the MD results and
show how does PF and MD compare. The VOF model and comparison between
the VOF and MD is shown in section~\ref{sec:VOF}. Next, in section~\ref{sec:discus}
we discuss the presented results and some open questions that remain. Finally,
in section~\ref{sec:concl} we draw conclusions from this study.

\section{Molecular dynamics simulations of water over silica-like substrate}
\label{sec:molecular-dynamics}

Molecular dynamics describes
the system on a molecular level. The simulated system consists of
water molecules, which interact
with other water molecules as in the system as well as the substrate through a specified force field.
Molecule positions and velocities are then
integrated over time and the results sampled
to obtain a representation of continuum variables.

\subsection{Setup}

\begin{figure}
    \centering
    \subfloat[]{\includegraphics{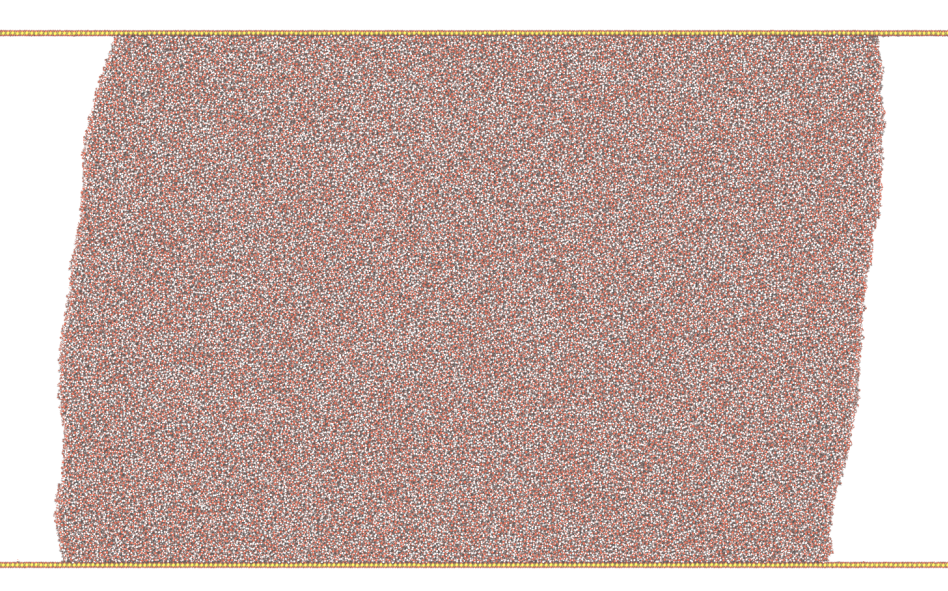}}
    \hspace*{10pt}
    \subfloat[]{\includegraphics{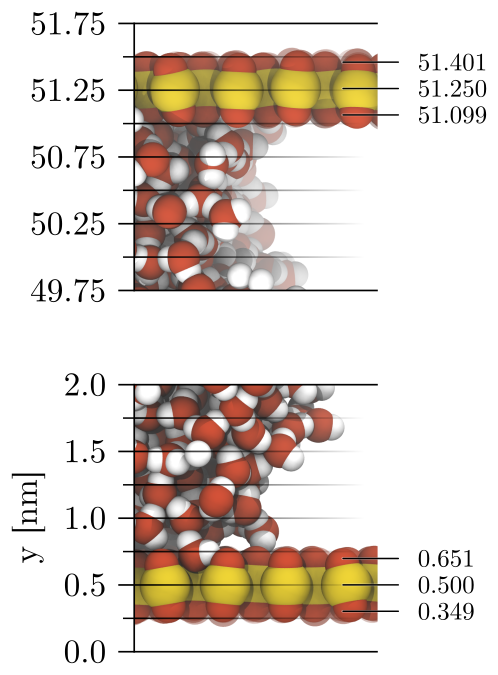}}
    \caption{\label{fig:md-system-setup} (a) Molecular view of the sheared
    water droplet configuration in steady regime. (b) Close up on upper and
    lower contact lines. The binning for measurements from MD
    along the $y$ axis is 0.25~nm and depicted with horizontal lines. The water slab begins in the fourth bins from the bottom and top.
    The coordinates of the wall atoms are marked on the right.}
\end{figure}

To represent the chosen water-vapour geometry (Fig.~\ref{fig:set-up}),
a two-dimensional shear system with water in vacuum is created by placing a water
slab between two walls, as shown in Fig.~\ref{fig:md-system-setup}(a).
After the initial equilibration step, some of the water molecules move to the void
and essentially form a very sparse water vapour.
The walls consist of rigid \SiO{} molecules which are neutral electric quadrupoles.
Water cannot easily slip over this substrate due to the electrostatic interactions
between water molecules, which are dipoles, and the substrate quadrupoles.
Water forms \emph{hydrogen bonds} with this substrate, a form of bond that is
transient but strong enough to prohibit slip \cite{johansson2018molecular}.
\rev{Due to the local nature of the hydrogen bonding, only one layer of the
solid substrate is sufficient to capture the physics of the moving
contact line.
The SPC/E model is the simplest possible choice of a MD model capable
of describing the hydrogen bonding.
A more complex water model would improve the agreement between 
the physical properties of MD and real water, but would not yield fundamentally
different results for contact line motion.}
The surface-water interaction is tuned to yield equilibrium contact
angle $\theta_0 = \staticContactAngle{}^{\circ}$.

Data is collected inside bins of size $0.25 \times 0.25 \, \textrm{nm}^2$ along $x$ and $y$. These bins form a regular grid covering the entire system. The binning is visualised along the $y$ axis in Fig.~\ref{fig:md-system-setup}(b). Average water molecule velocity, mass and temperature is sampled inside all bins over a period of 50~ps, after which the data is stored and the bins are reset for the next sampling period.

We perform four simulations from different starting configurations. These configurations are created by generating the initial velocity field with different seeds for the random generator. Before the shear is applied each configuration is allowed to relax over a period of 100~ps.
\rev{Evaporation and condensation process is a liquid surface effect and we have
observed that 100~ps is sufficient to reach equilibrium vapour density.}
More details about the procedure are available in appendix~\ref{app:molecular-dynamics}.

\subsection{System parameters}
\label{sec:md-system-parameters}

Parameters for the MD system are reported in Fig.~\ref{fig:set-up}. Bulk water density $\rho_w$, viscosity $\mu_w$ and surface tension $\sigma$ are measured in separate simulations of pure water. Following Quan et al. \cite{qian2003molecular} the water slip over the substrate is characterised through a friction parameter $\beta_f = \mu_w / l_s$, where $l_s$ is the corresponding Navier slip length. The friction parameter $\beta_f = 5.9 \, \mu_w \, \textrm{nm}^{-1}$ is measured in a Couette flow setup using the boundary layer of water molecules in direct contact with the wall molecules. From this measurement, the Navier slip length $l_s = 0.17 \, \textrm{nm}$ is extracted. Finally, the interface width $\epsilon$ is taken as the length over which the density goes from bulk to vapour density. This change occurs over three bins, giving $\epsilon = 0.75 \, \textrm{nm}$.

\subsection{Results}

\begin{figure}
    \centering
    \subfloat[]{\includegraphics[height=0.495\linewidth]{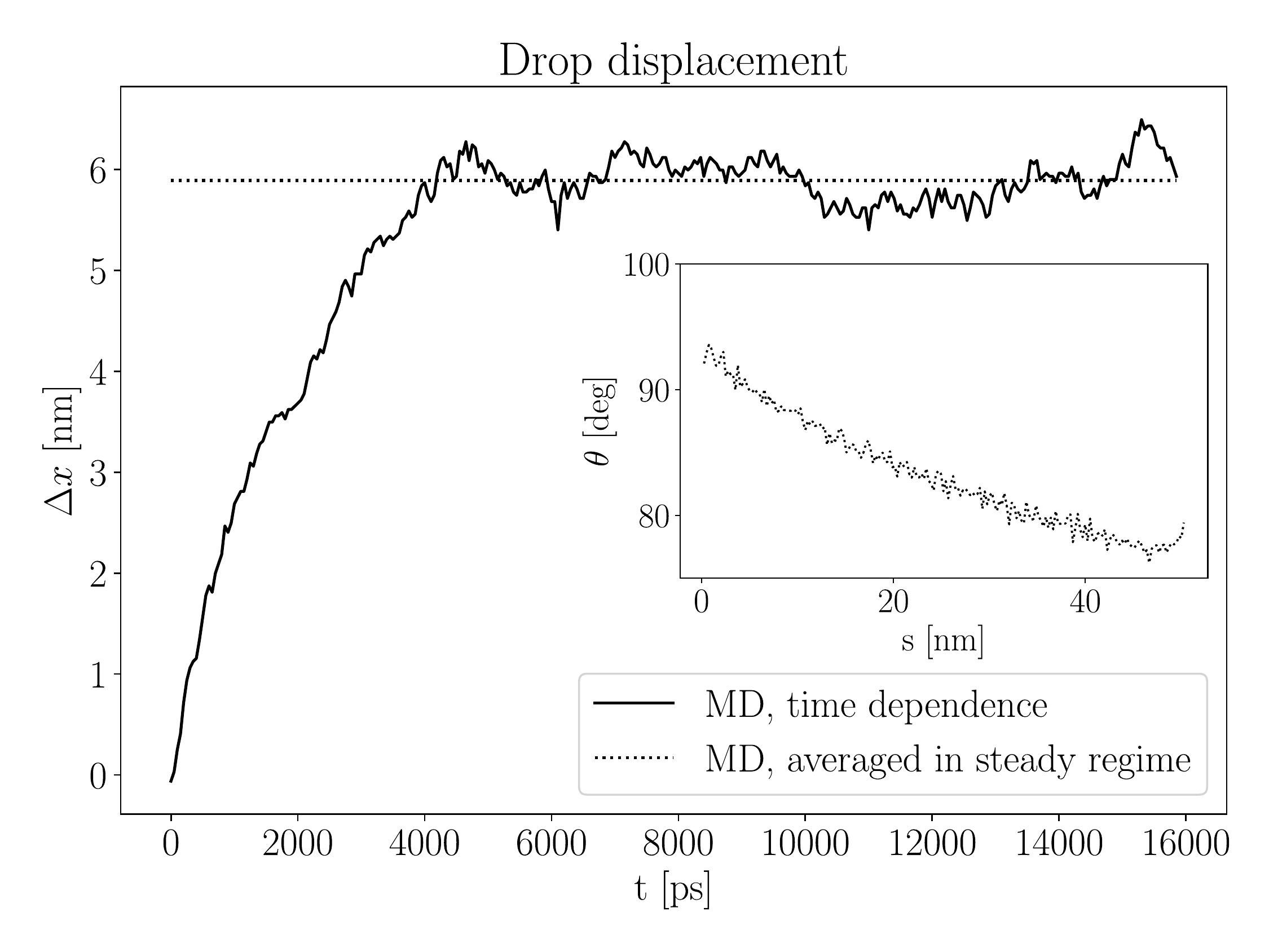}}
    \subfloat[]{\includegraphics[height=0.495\linewidth]{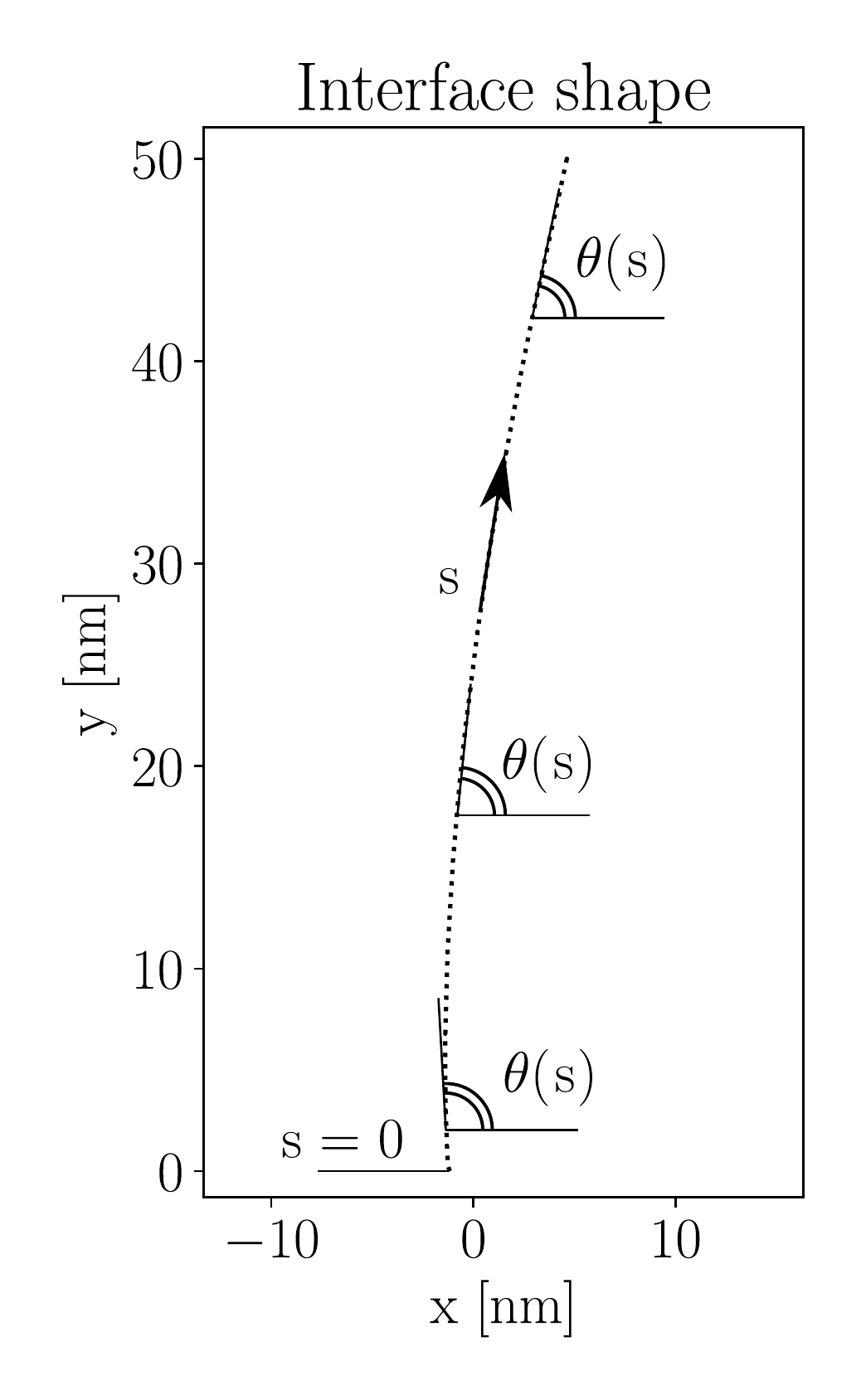}}
    \caption{\label{fig:md-results} (a) Shear separation $\Delta x(t)$ from the four MD simulations with mean value $\Delta x = 5.89$ nm in the steady regime. (b) Average interface shape in the steady state. The average is done around the midpoint, at $y = 25 \, \textrm{nm}$. As additional measure, we extract the interface
    angle along the curvilinear coordinate $s$ as illustrated in (b). The interface
    angle from MD is shown in the inset of (a).}
\end{figure}

Shear experiments consist of two stages. Once the shear is applied the water slab deforms until reaching its equilibrium (steady state). For the transient state we measure the separation $\Delta x(t)$ between the upper and lower contact line positions, which begins at zero in the original slab geometry and increases to $\Delta x_s \left(t\right)$ in
the steady state.
To identify the the interface between the water phase and vapour phase, we locate the
bins with a water density measurement, which is half of the bulk density.
The mean (averaged between four different runs) separation for the four simulations
is shown in Fig.~\ref{fig:md-results}(a) with solid line. The steady state is
reached in 5.9~ns, after which the data has been averaged in time to obtain the
final mean displacement 
$\Delta x_m = 5.89$~nm.

In the steady state the slabs are characterised using the full interface shape $x(y)$ and the flow field. We mirror the right interface along both axes and overlay it on top of the left interface. Thus the lower end represents the \emph{receding} contact line and the upper end the \emph{advancing}. The average of all interfaces from MD in the steady
regime is shown in Fig.~\ref{fig:md-results}(b). To gain more detailed insight into the
agreement between MD and continuum models, we introduce parametric coordinate $s$ 
as illustrated in Fig.~\ref{fig:md-results}(b). The interface angle
$\theta\left(s\right)$ is shown in the inset of Fig.~\ref{fig:md-results}(a).
Finally, we extract an averaged flow field from the MD runs, which we will use to discuss the agreement between MD, PF and VOF (Figs.~\ref{fig:try-perfect-match-itf}, \ref{fig:try-perfect-match-flow}, \ref{fig:relax-angle-fit}, \ref{fig:Nav-slip-dif-fits} and \ref{fig:vel-near-CL}).

\section{Phase-field model} \label{sec:pf}

We consider a 2D phase-field (PF) model of a binary mixture to model two regions of different densities and viscosities. The water and vapour phases in Fig.~\ref{fig:set-up} are assumed to be incompressible and the interface between the regions to be diffuse, i.e. that quantities vary smoothly over the interface.

\subsection{Governing equations for the binary mixture}

The phase-field model introduces a phase variable $C(x,y)$
ranging from $1$ in the water phase to $-1$ in the vapour.
The derivation of the governing equations for the phase variable
can be found in \cite{carlson2012thesis,jacqmin2000contact}.
The phase-field variable is governed by a convection-diffusion equation
in a yet undetermined flow field $\vec{u}$ as
\begin{equation}
\frac{\pd C}{\pd t} = \nabla \cdot \left[ M \nabla
\left( \beta \Psi'\left(C \right) - \alpha \nabla^2 C \right) \right] - \vec{u} \cdot \nabla C \, .
\label{eq:cahn-hil-dim}
\end{equation}
The diffusive flux  $\vec{J}_d = - M \nabla \phi$ is proportional to the gradient of the chemical
potential, defined as
$\phi = \beta \Psi'\left(C \right) - \alpha \nabla^2 C$,
 with proportionality coefficient $M$, which is called
the phase-field mobility. Due to the assumption of an incompressible flow, 
the convective flux takes the simple form
$\vec{J}_c = \vec{u} \cdot \nabla C$.
In the chemical potential, we have two parameters $\alpha$ and $\beta$, which are
related to the surface tension as $\sigma = 2 \sqrt{2 \alpha \beta} /3$ and the
characteristic thickness
of the diffuse interface as $\epsilon = \sqrt{\alpha/\beta}$. In addition, it
contains the derivative of the standard double-well
potential $\Psi \left(C\right) = \left( C+1 \right)^2\left( C-1 \right)^2/4$.

The motion of the fluids is described by the incompressible
Navier--Stokes equations with variable density and viscosity. We define the density
and viscosity as
\begin{equation}
\rho\left(C\right) = \rho_w \frac{C+1}{2} - \rho_g
\frac{C-1}{2} \ \ \ \ \mbox{and} \ \ \ \
\mu\left(C\right) = \mu_w \frac{C+1}{2} - \mu_g
\frac{C-1}{2}, \label{eq:pf-def-den-visc}
\end{equation}
respectively. Here $\rho_w$ and $\mu_w$ are the density and the viscosity of the water,
and $\rho_g$ and $\mu_g$ are the density and the viscosity of the vapour
component. The Navier--Stokes equations then become
\begin{align}
\rho \left( C \right) \left[ \frac{\pd \vec{u}}{\pd t} + \left(\vec{u} \cdot \nabla
 \right) \vec{u} \right] & = - \nabla P + \nabla \cdot \left[
 \mu \left( C \right) \left\{ \nabla \vec{u} + \left( \nabla \vec{u} \right)^T \right\}
 \right] + \vec{f}_\sigma, \label{eq:ns-1}\\
 \nabla \cdot \vec{u} & = 0, \label{eq:ns-2}
\end{align}
where the volume force $\vec{f}_\sigma = - C\, \nabla \phi$ corresponds
to the surface tension
force and acts over the diffuse interface region. This form of the surface tension
forcing is the so called potential form~\cite{jacqmin1999calculation}, which
uses a reduced pressure.

\subsection{Boundary conditions for the phase-field model}

The convection-diffusion equation (\ref{eq:cahn-hil-dim}) is a fourth-order
partial differential equation and requires two boundary conditions. First, we impose a non-equilibrium wetting condition~\cite{jacqmin2000contact,qian2003molecular,carlson2011dissipation} on
the solid moving wall,
\begin{equation}
- \mu_f \epsilon \left( \frac{\pd C}{\pd t} + \vec{u} \cdot \nabla C \right) =
\alpha \nabla C \cdot \hat{n} - \sigma \cos \left( \theta_0 \right) g'\left( C \right),
\end{equation}
where $\mu_f$ is a contact line friction parameter, having the same units
as bulk dynamic viscosity.
Here, $\theta_0$ is the equilibrium contact angle
and $g\left(C\right) = 0.5 - 0.75 C + 0.25 C^3$ is a switching
function describing a smooth transition from water
to vapour.
The unit normal vector $\hat{n}$ is directed from the fluid to the surrounding
solid, which is standard definition in numerical approaches.
%
%
If one sets $\mu_f = 0$, the dynamic contact angle is always
enforced to the equilibrium angle $\theta_0$.
Non-zero contact line friction allows the dynamic contact angle to evolve naturally as
a function of contact line speed.
The second boundary
condition for the phase function is zero
diffusive flux of chemical potential through the boundaries, i.e. $\nabla \phi \cdot
\hat{n} = 0$. On the outer sides of the domain, periodic boundary
conditions are enforced.


The fluid momentum equations are subject to zero wall-normal velocity,
$u_y = 0$.  
For the tangential velocity component, we
first consider the classical no-slip condition, which is equivalent to setting
the fluid velocity near the moving wall to the velocity of the wall, $u_x = U_w$.
In this situation, the only mechanism through which the contact line can move is
the diffusion of the phase-field variable.
We also impose the Navier slip
condition, $u_x = U_w - l_s\,\pd_y u_x\,\hat{n}_y$,
where $l_s$ is the Navier slip length.
%
%
The final boundary condition which we investigate for the PF model is the
generalised Navier boundary condition (GNBC). This boundary
condition (using the friction factor $\beta_f$
for the slip velocity \cite{qian2003molecular}) takes the form
\begin{equation}
\beta_f \left( u_x - U_w \right) = - \mu\,\pd_y u_x\,\hat{n}_y
+ \left[ \alpha\, \pd_y C \, \hat{n}_y - \sigma \cos \left( \theta_0 \right) g'\left( C \right) \right] \pd_x C \label{eq:pf-ns-bc3}
\end{equation}
where the second term is the uncompensated Young's stress.
We assume a constant slip length $l_s$
over the whole solid surface. This leads to a friction coefficient $\beta_f$, which
varies in the same way as the liquid viscosity.
The equations are implemented in a finite-element solver (for more details
see appendix~\ref{app:pf}).

\subsection{Comparison with molecular dynamics}

The PF parameters contain quantities
uniquely determined from MD 
($\rho_w$, $\mu_w$, $\sigma$, $\theta_0$, $\epsilon$) 
and quantities that are
less known or fully unknown 
($\mu_f$, $M$, $\rho_g$ and $\mu_g$). The vapour properties are not known
due to the small
system size, which renders any measurements from the MD outside of the liquid
phase impractical. Therefore we have fixed both $\rho_g$ and $\mu_g$ in the
PF (and also later in the VOF) simulations as small as
numerically feasible, $\rho_g = 10^{-3}\,\rho_w$ and
$\mu_g = 10^{-3}\,\mu_w$, see also Fig.~\ref{fig:set-up}.
\rev{The limiting factor for our implementation is viscosity ratio.
The small density and viscosity in vapour phase is motivated by the
fact that in MD there are only a handful of vapour molecules outside
the water droplet (appendix~\ref{app:molecular-dynamics}), and those consequently
does not exhibit any notable stress on the liquid phase.
An alternative approach could be to read off the density and viscosity data
from measurements \cite{engtoolb_watvapor2004,engtoolb_gasviscos2014}, which
would yield $\rho_g = 2.64 \cdot 10^{-5}\,\rho_w$ and
$\mu_g = 1.13 \cdot 10^{-2}\,\mu_w$. We have
checked that using density and viscosity from literature for PF simulations
yields only minor changes and does not affect the conclusions of this study.}
The remaining unknown parameters ($\mu_f$ and $M$) are fitted to the MD data.

\begin{figure}[ht!]
\centering
\subfloat[]{\includegraphics[height=6.4cm]{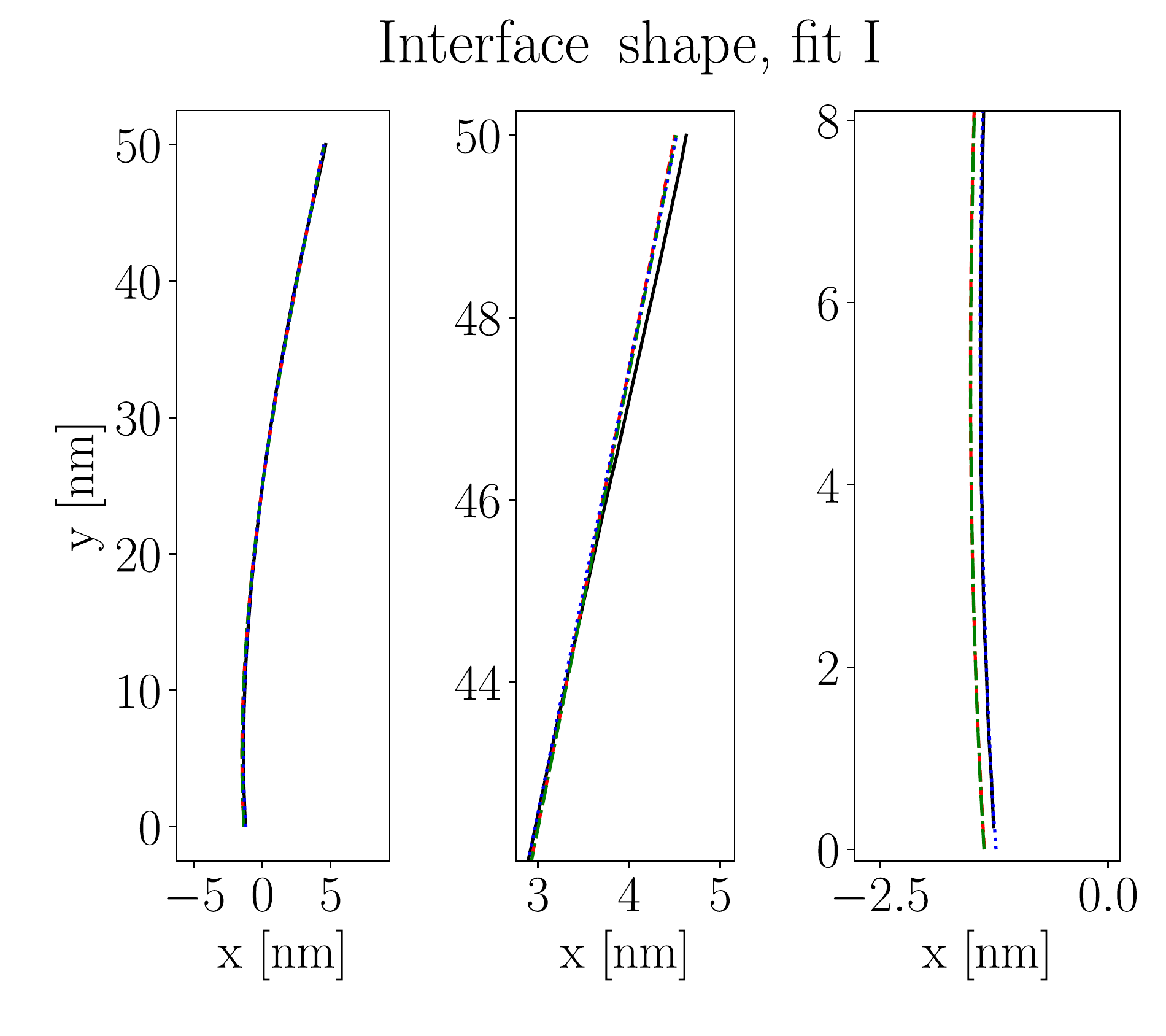}}
\subfloat[]{\includegraphics[height=6.4cm]{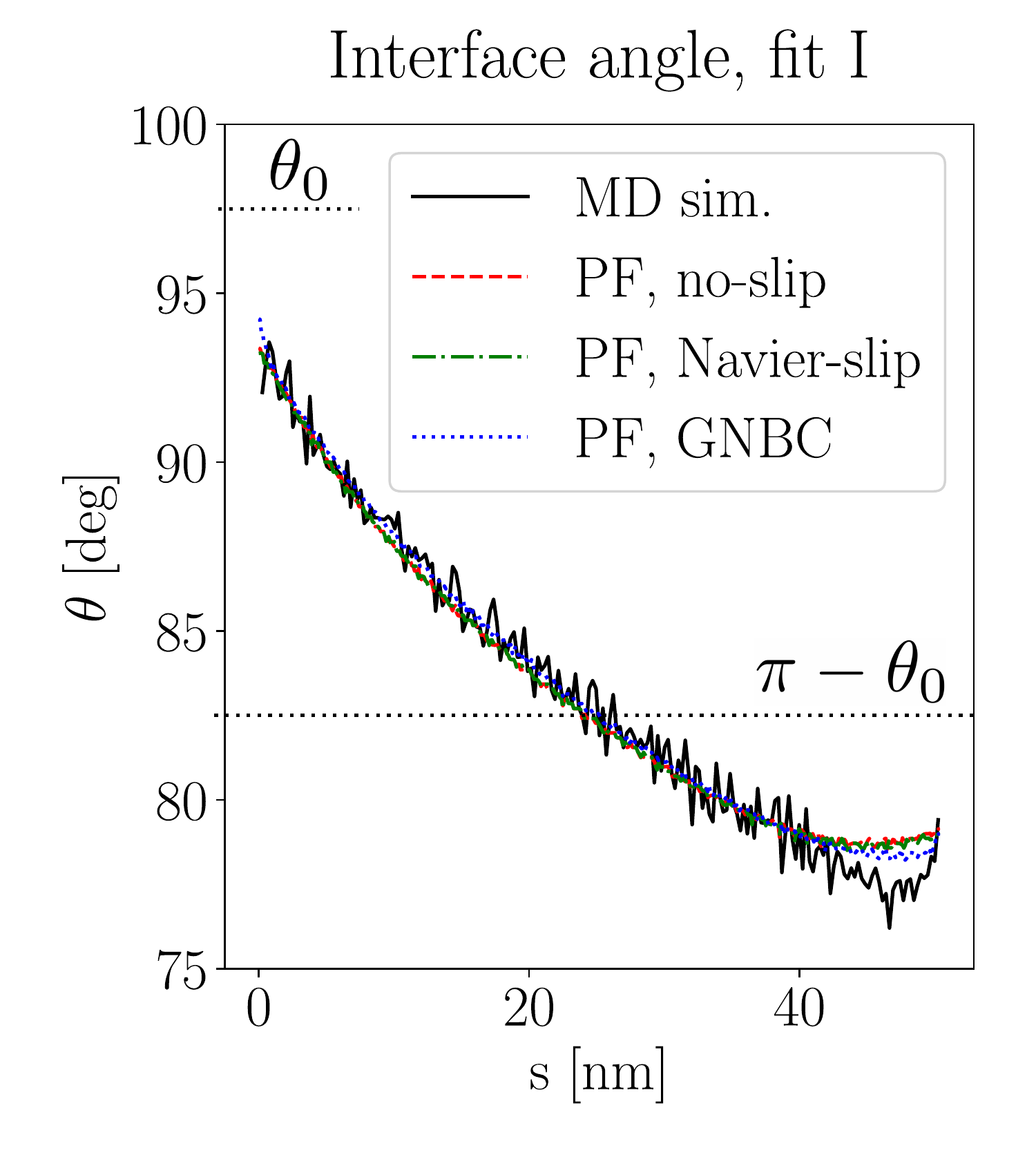}}
\caption{Results of the fitting procedure I (accurate dynamic
contact angle). Steady interface shape (a)
and angle distribution over the interface (b) from MD
simulations and fitted PF simulations with different boundary conditions.
Static contact angles for advancing and receding contact lines are
indicated with dotted black lines.}
\label{fig:try-perfect-match-itf}
\end{figure}

The fitting procedure I, which we 
later refer to as  ``fit I'', is as follows:
\begin{enumerate}
  \item adjust the contact line friction ($\mu_f$) individually for advancing
  and receding contact lines to match the MD dynamic contact angles;
  \item adjust the phase-field mobility ($M$) to match the MD drop displacement 
  $\Delta x_m$.
\end{enumerate}
Results are shown in Fig.~\ref{fig:try-perfect-match-itf}(a,b).
There is a local error in the interface angle (Fig.~\ref{fig:try-perfect-match-itf}b)
near the advancing contact line.
The drop displacement from PF agrees with $\Delta x_m$ with an accuracy
of $2\%$,
which
we consider a very good match. The parameters needed to arrive with the corresponding
PF results along with the
obtained drop displacement
are listed in Tab.~\ref{tab:fullFit-param-summary}.


\begin{table}[ht!]
\begin{center}
\begin{tabular}{p{30mm}|p{20mm}|p{25mm}|p{20mm}|p{15mm}}
 & PF no-slip & PF Navier-slip & PF GNBC & MD \\ \hline
$\mu_{fa}/\mu_w$ & $2.5$ & $2.8$ & $12$ & -\\
$\mu_{fr}/\mu_w$ & $3.0$ & $3.3$ & $13$ & - \\
$\Pe{} = U \epsilon L / \left( M \sigma \right)$ & $0.070$ & $0.070$ & $0.075$ & - \\
$\Delta x$ & $5.85$ nm & $5.87$ nm & $5.75$ nm & $5.89$ nm \\
\end{tabular}
\end{center}
\caption{Summary of the obtained phase-field parameters and displacement, fit I.}
\label{tab:fullFit-param-summary}
\end{table}

\begin{figure}[ht!]
\centering
\subfloat[MD adv.]{\includegraphics[height=2.0cm]{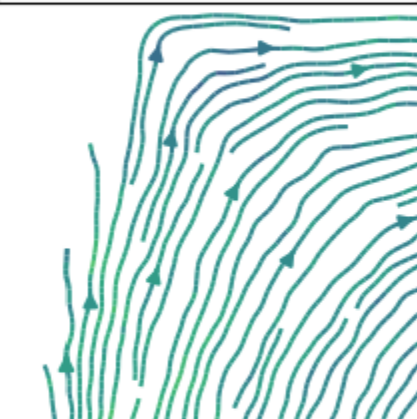}}
\hspace*{5pt}
\subfloat[PF GNBC adv.]{\includegraphics[height=2.0cm]{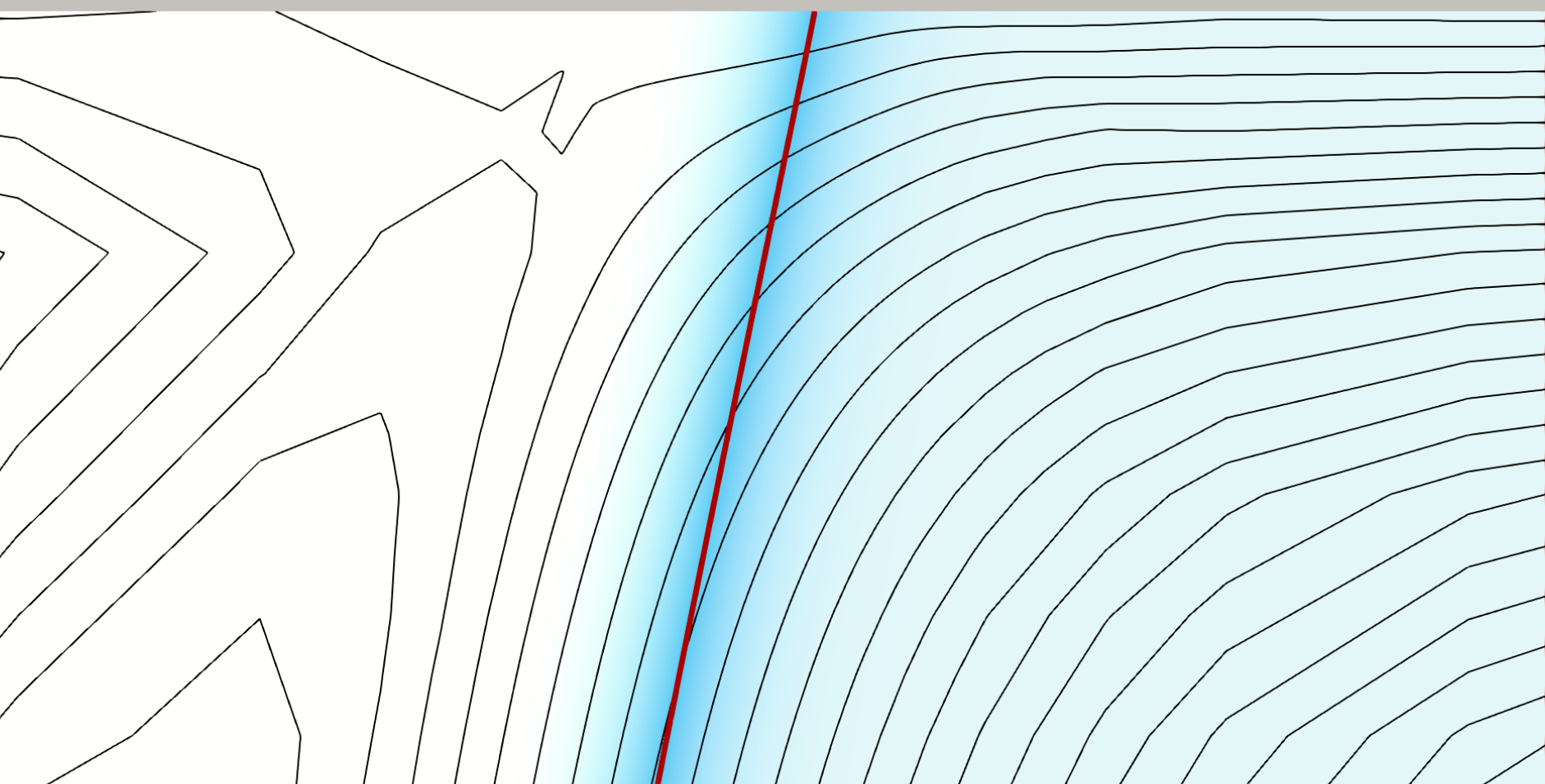}} 
\hspace*{5pt}
\subfloat[MD rec.]{\includegraphics[height=2.0cm]{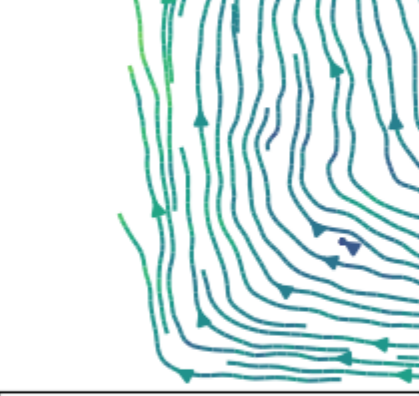}}
\hspace*{5pt} 
\subfloat[PF GNBC rec.]{\includegraphics[height=2.0cm]{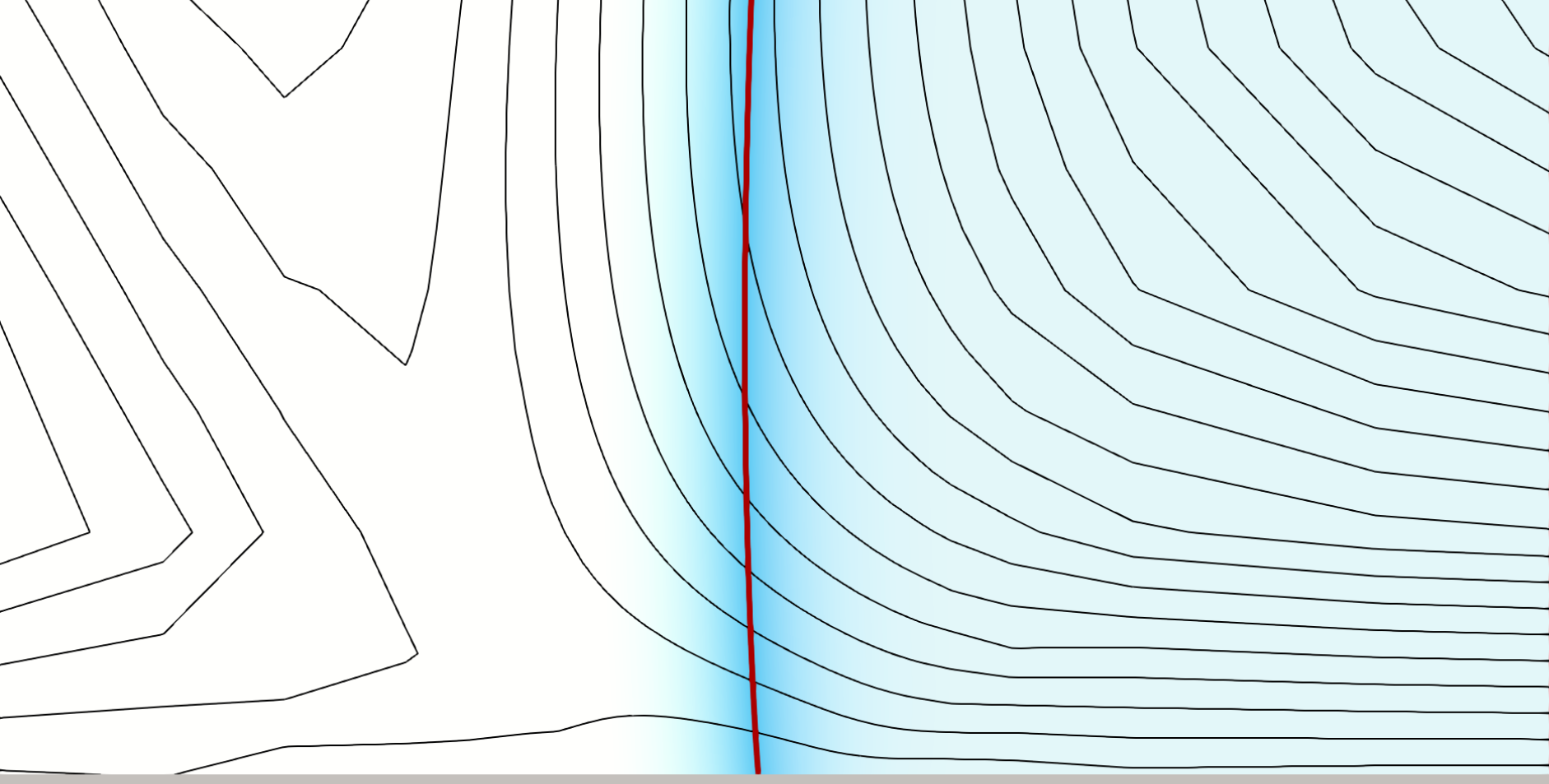}}
\caption{Flow field results of the fitting procedure I.
Streamlines
near advancing contact line (a,b) and near receding contact line (c,d) from MD
simulations (a,c) over a $10$ nm $\times 10$ nm patch and PF simulations
with GNBC boundary condition (b,d) over $20$ nm $\times 10$ nm patch. With red
line, we indicate the isoline of $C = 0$, which represents the interface.
The light blue background field show the variation of the phase-field
variable $C$.}
\label{fig:try-perfect-match-flow}
\end{figure}

A 10~nm~$\times$~10~nm close-ups of MD streamlines
near the advancing and the receding contact lines are shown in
Fig.~\ref{fig:try-perfect-match-flow}(a,c), respectively. For comparison,
in Fig.~\ref{fig:try-perfect-match-flow}(b,d) we show
20~nm~$\times$~10~nm close-ups of streamlines from PF with GNBC.
We observe that PF streamlines cross the interface and
extend several nanometers into the vapour. This is in contrast to the MD results
and therefore the PF predictions obtained through fitting procedure I 
provide un-physical flow fields.
The PF flow with no-slip and Navier-slip conditions show slightly worse
agreement and are not reported. The extent of which PF streamlines
cross from water to vapour in the PF results can be characterised with the \Pe{} 
number (appendix~\ref{app:pf}), quantifying the relative importance between convection and diffusion and defined as
$\Pe{} = U \epsilon L / \left( M \sigma \right)$. From the last row in Tab.~\ref{tab:fullFit-param-summary}
we observe that all fits have resulted in a \Pe{} which is much smaller than unity: 
a high diffusion regime, in contrast to the convection dominated
MD data. The high diffusion leads to the many streamlines crossing the PF interface.
However, if one would use a much larger \Pe{} number (much smaller mobility) but keep other 
parameters the same, the MD steady state drop displacement $\Delta x_m$ would be largely overestimated.

To obtain a more physical flow field, we devise another fitting procedure,
in which we relax the requirement on the dynamic contact angle.
Fitting procedure II (``fit II'') is defined as:
\begin{enumerate}
  \item for PF simulation with no-slip boundary conditions,
  we select $\mu_f = \mu_w$
  and then vary $M$ to match the drop displacement $\Delta x_m$ observed in MD;
  \item for PF simulations with Navier-slip and GNBC boundary conditions,
  we adjust $\mu_f$ to
  match the contact angle in the final no-slip simulation, carried out
  in point 1. of this strategy, and then fit $M$ to match
  the drop displacement $\Delta x_m$ from the MD.
\end{enumerate}
\begin{figure}[ht!]
\centering
\begin{tabular}[c]{ccc}
\multirow{2}{*}[5pt]{ 
\subfloat[]{\includegraphics[height=4.8cm]{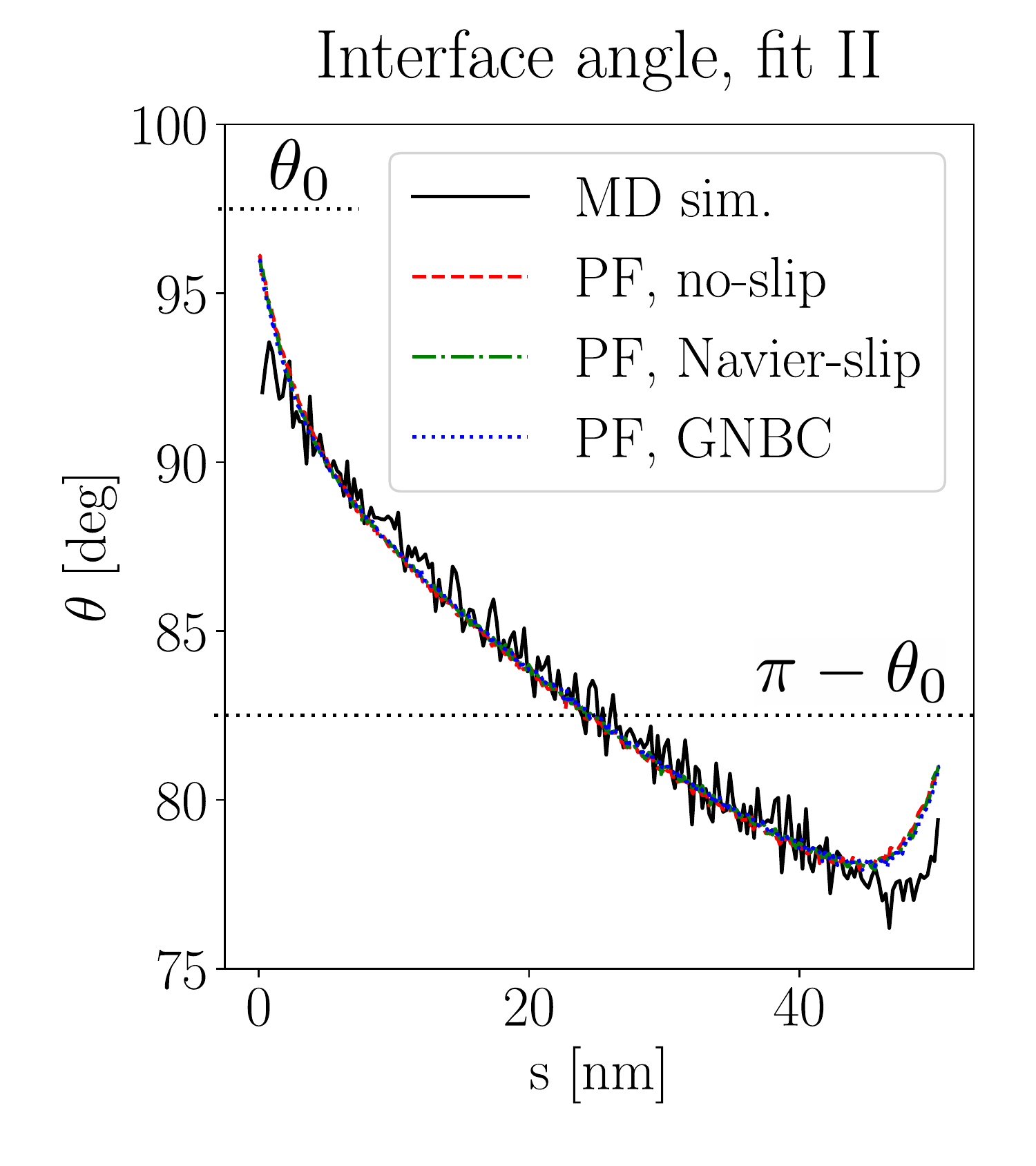}}
\hspace*{-10pt}
} &
\subfloat[MD]{\includegraphics[height=1.8cm]{MD_streamlines_advancing_10x10nm}} &
\subfloat[PF No-slip]{\includegraphics[height=1.8cm]{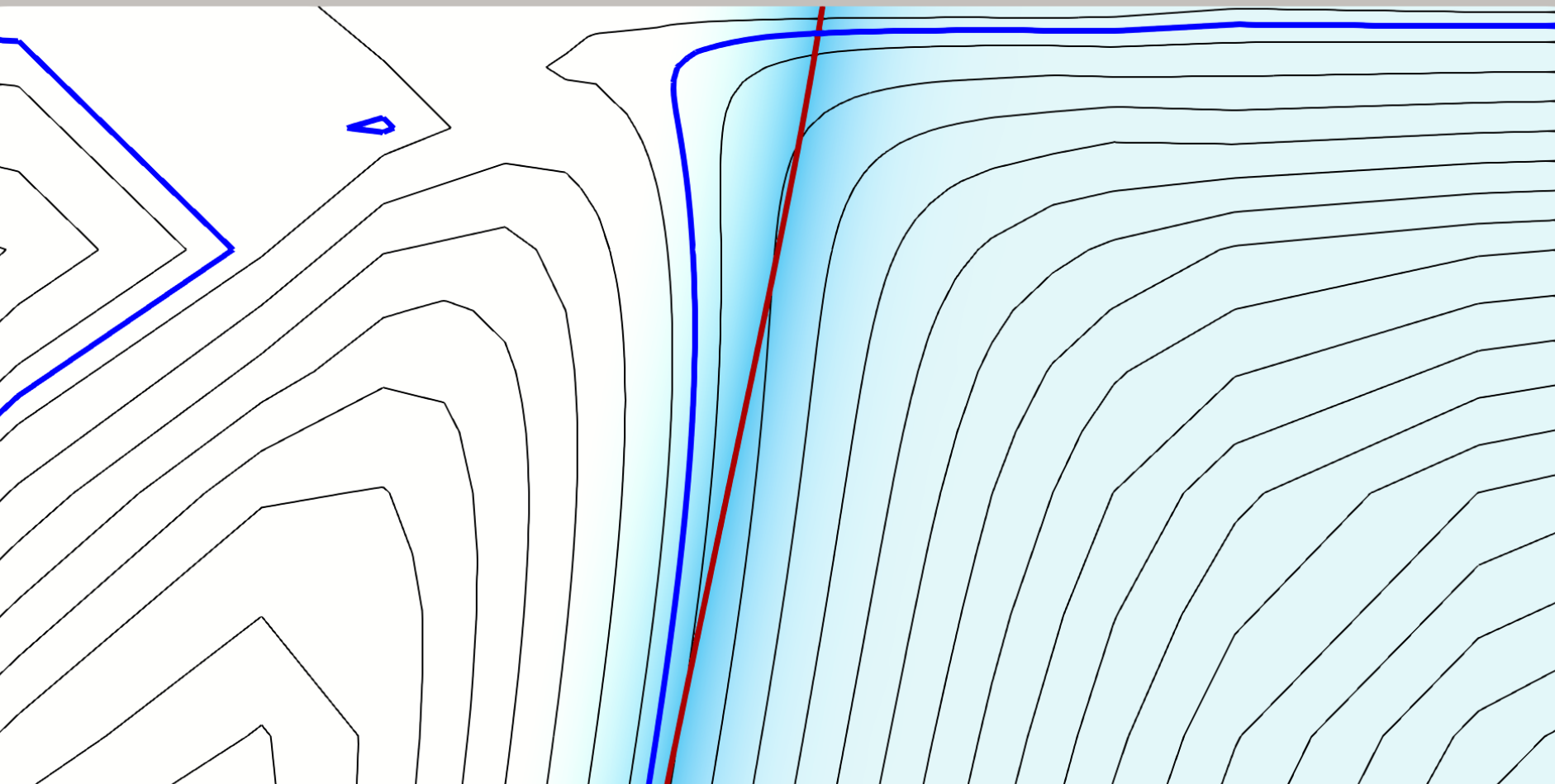}} \\
 &
\subfloat[PF Navier-slip]{\includegraphics[height=1.8cm]{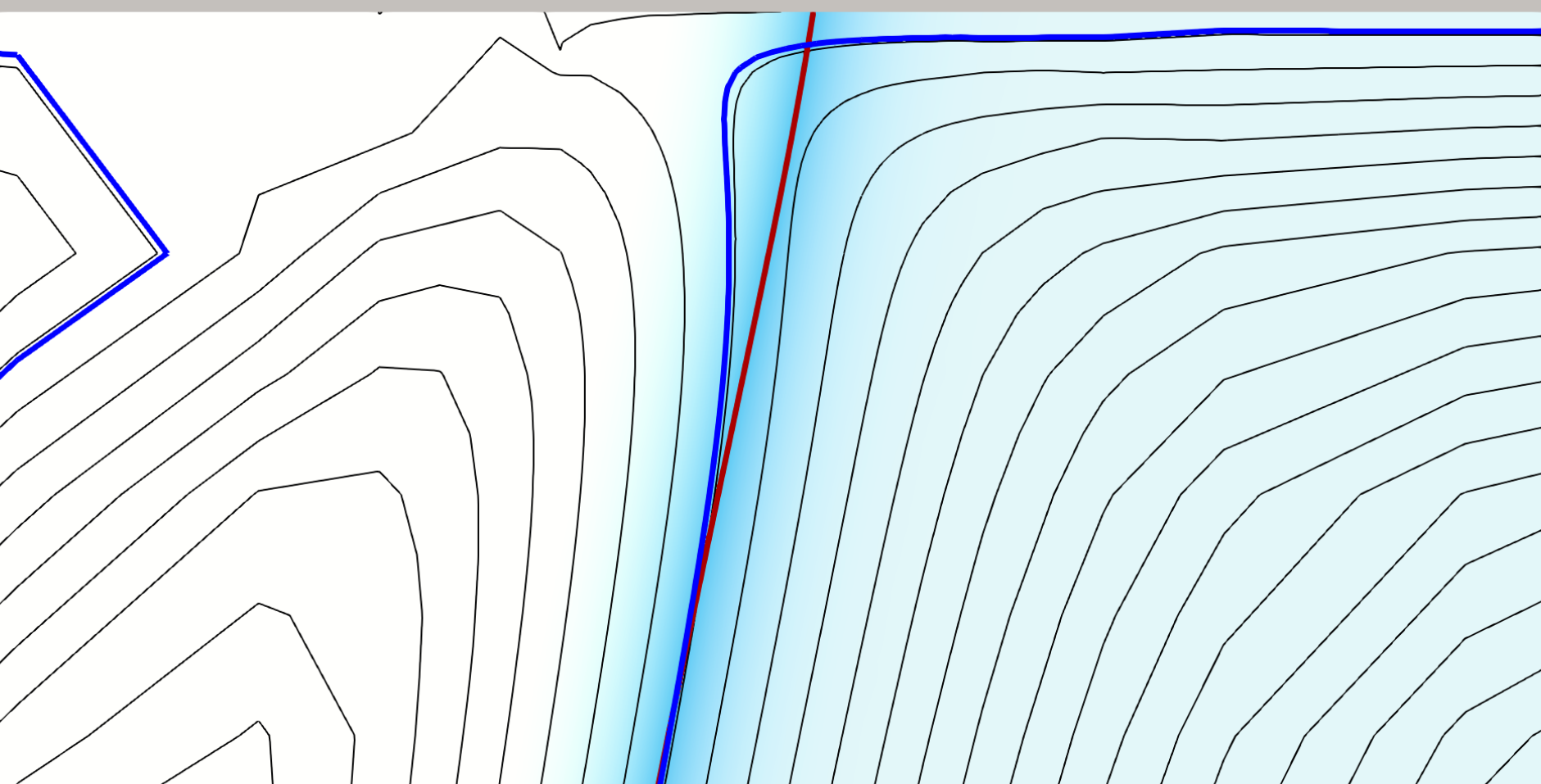}}
\hspace*{5pt} &
\subfloat[PF GNBC]{\includegraphics[height=1.8cm]{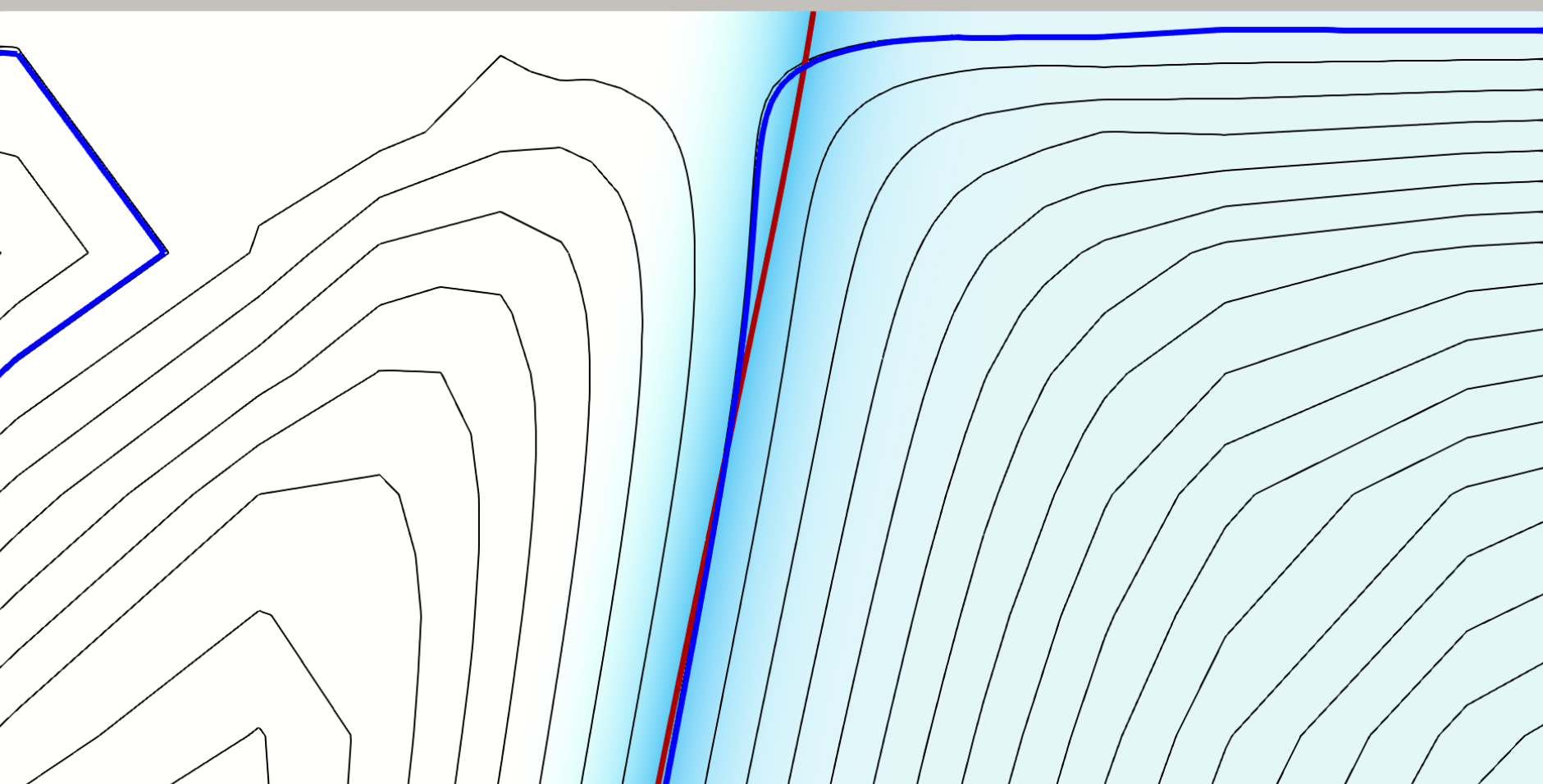}}
\hspace*{5pt} \\
\end{tabular}
\caption{Results of the fitting procedure II (larger error on dynamic
contact angle). Angle distribution over
the interface (a) from MD
simulations and fitted PF simulations with different boundary conditions. Streamlines
near advancing contact line (b-e) from MD
simulations (b) over a $10$ nm $\times 10$ nm patch and PF simulations
with different boundary conditions (c-e) over $20$ nm $\times 10$ nm patch. With red
line, we indicate the isoline of $C = 0$, which represents the interface.
The dark blue line represents a streamline originating 0.25 nm away
from the wall within the liquid drop.}
\label{fig:relax-angle-fit}
\end{figure}
\begin{table}[ht!]
\begin{center}
\begin{tabular}{p{30mm}|p{20mm}|p{25mm}|p{20mm}|p{15mm}}
 & PF no-slip & PF Navier-slip & PF GNBC & MD \\ \hline
$\mu_{fa}/\mu_w$ & $1.0$ & $1.1$ & $2.0$ & - \\
$\mu_{fr}/\mu_w$ & $1.0$ & $1.1$ & $2.0$ & -\\
$\Pe{} = U \epsilon L / \left( M \sigma \right)$ & $0.8$ & $1.4$ & $1.8$ & - \\
$\Delta x$ & $5.84$ nm & $5.82$ nm & $5.81$ nm & $5.89$ nm \\
\end{tabular}
\end{center}
\caption{Summary of the obtained phase-field parameters and displacement, fit II.}
\label{tab:relaxAngle-param-summary}
\end{table}
We carry out these fits for all considered boundary conditions. By investigating
the obtained interface angle distribution (Fig.~\ref{fig:relax-angle-fit},a) we see that
overall agreement is good, while the local error near contact lines is increased. This
difference is, however, not visible in the interface shape: practically the same
agreement as presented in Fig.~\ref{fig:try-perfect-match-itf}(a) is obtained.
However, investigating
the flow field near the advancing contact line reveals much better agreement with the
MD results, although there is a small overshoot through the interface of streamlines
close to the wall (for fair measurement of the overshoot, we have identified
a single streamline in all the simulations, which originates 0.25~nm away from the
wall within the drop). The improved agreement with the MD (and smaller overshoot of the
streamlines) is due to a smaller phase-field mobility (or larger \Pe{} number, see
Tab.~\ref{tab:relaxAngle-param-summary}). The reduced contact line friction
is the reason why it is possible to use smaller
phase-field mobility. Reducing contact line friction leads to smaller displacement of the
drop, which consequently allows to reduce the phase-field mobility to increase the
drop displacement back to the $\Delta x_m$.

\begin{figure}[ht!]
\centering
\begin{tabular}[c]{ccc}
\multirow{2}{*}[5pt]{ 
\subfloat[]{\includegraphics[height=4.8cm]{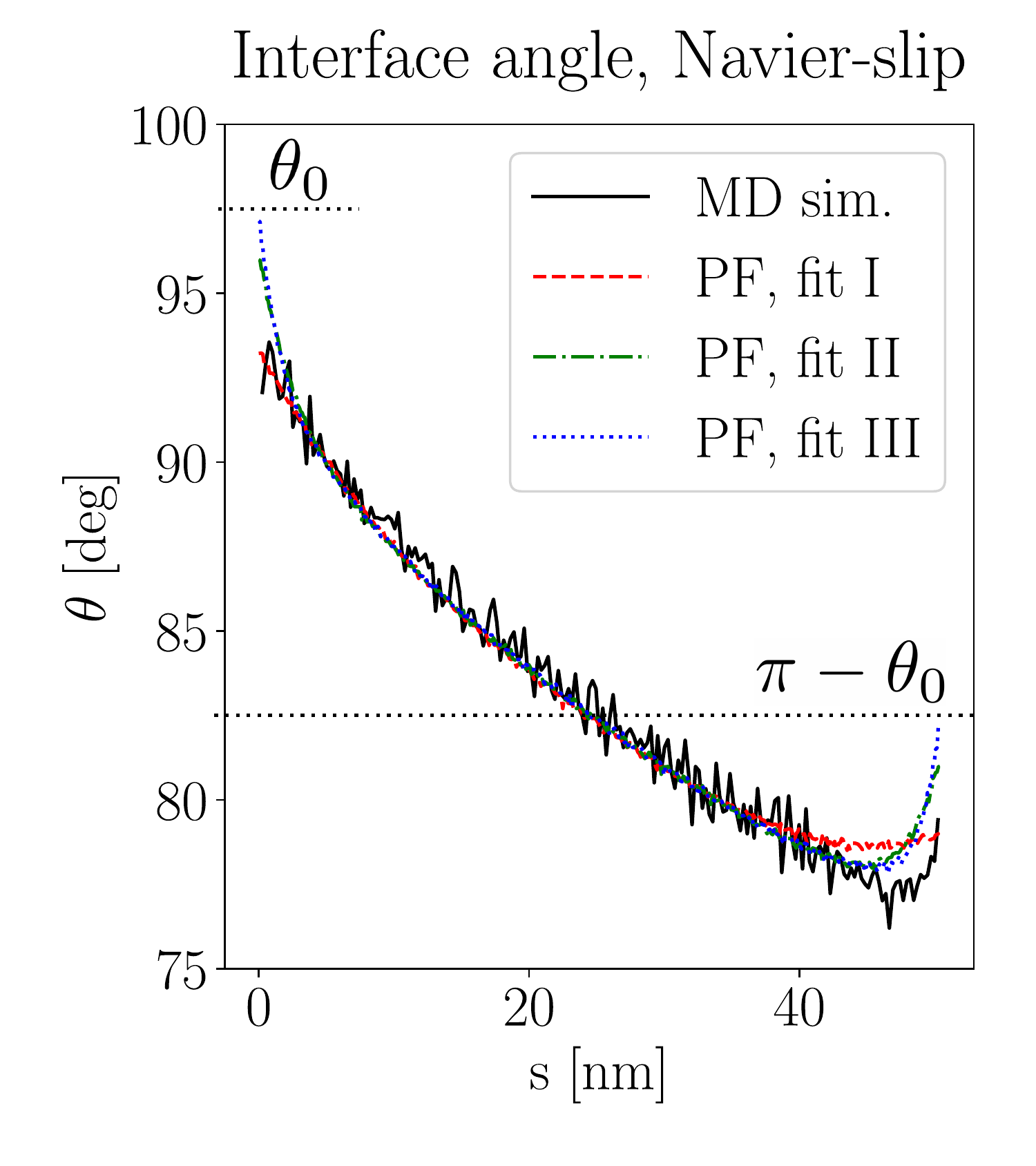}}
\hspace*{-10pt}
} &
\subfloat[MD]{\includegraphics[height=1.8cm]{MD_streamlines_advancing_10x10nm}} &
\subfloat[PF Navier-slip fit I]{\includegraphics[height=1.8cm]{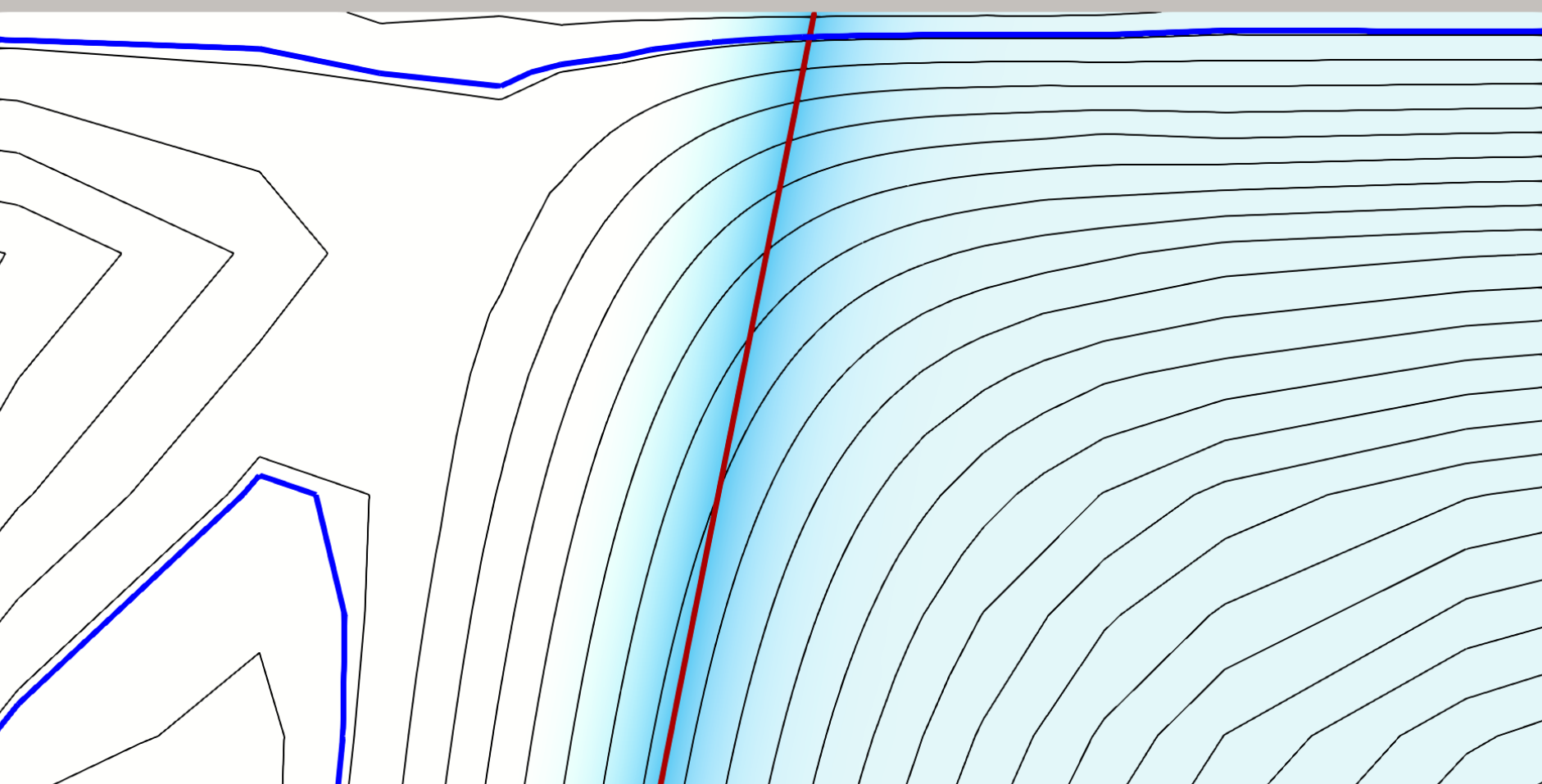}} \\
 &
\subfloat[PF Navier-slip fit II]{\includegraphics[height=1.8cm]{PF_relaxAngleSame_streamlines_advancing_10x20nm_NavSlip_new}}
\hspace*{5pt} &
\subfloat[PF Navier-slip fit III]{\includegraphics[height=1.8cm]{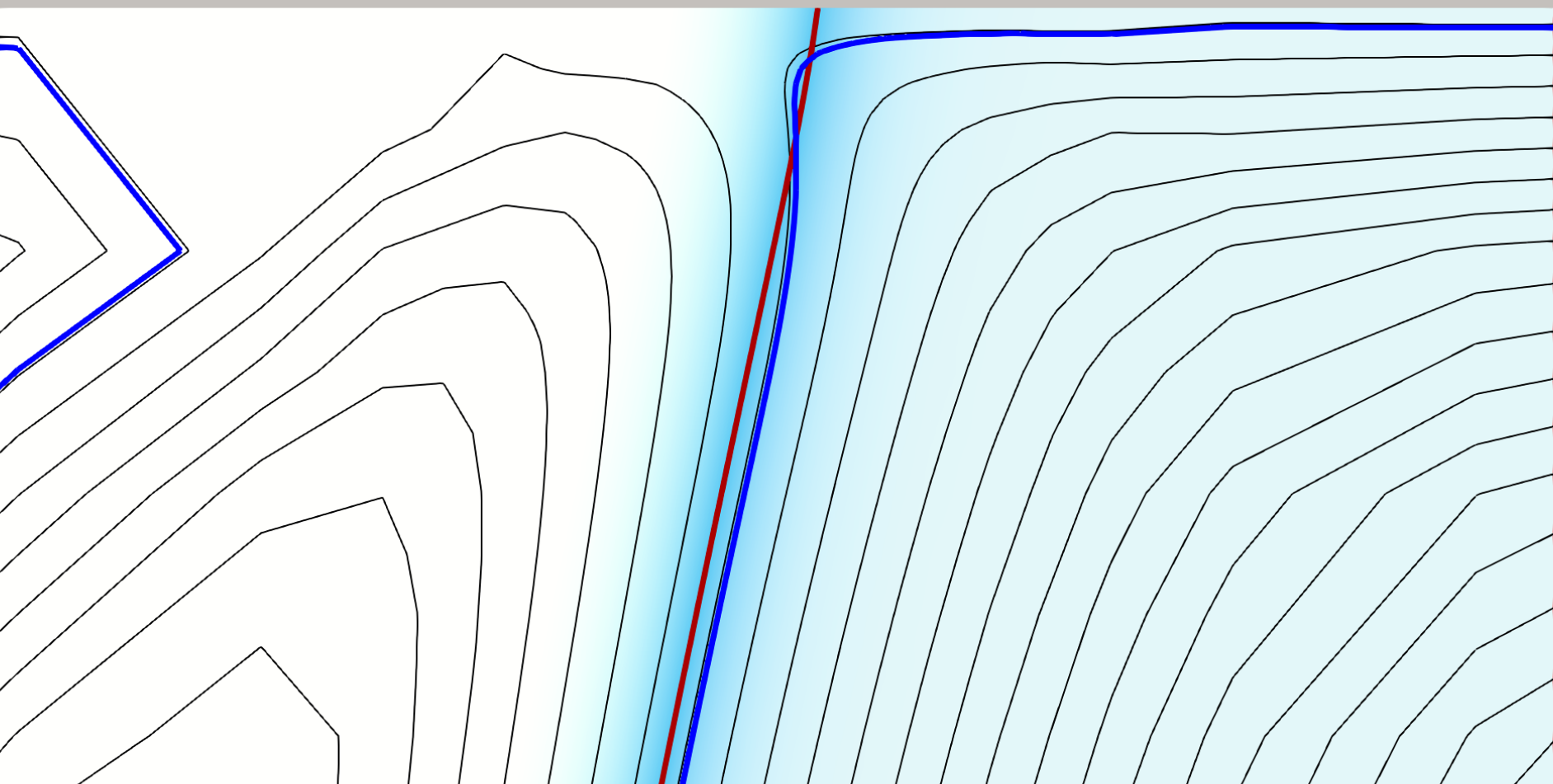}}
\hspace*{5pt} \\
\end{tabular}
\caption{Compilation of PF simulation results with Navier-slip
boundary condition, including results from fit III. Same measures
are reported as in Fig.~\ref{fig:relax-angle-fit}.
}
\label{fig:Nav-slip-dif-fits}
\end{figure}

These results suggest that as the
phase-field mobility $M$ is reduced (or \Pe{} number is increased),
the flow field near contact line approaches the one observed in MD. Therefore
we devise the final fitting procedure III (``fit III''), which is defined as:
\begin{enumerate}
  \item set the contact line friction $\mu_f = 0$ (fixing the dynamic
  contact angle to the equilibrium one). This gives minimum friction at the
  contact line;
  \item fit the phase-field mobility $M$ to obtain the drop displacement $\Delta x_m$.
\end{enumerate}
The obtained \Pe{} numbers using fitting procedure III are given in Tab.~\ref{tab:maxPe-param-summary}.
The results from PF simulations with Navier-slip boundary condition 
from all fitting procedures are shown
in Fig.~\ref{fig:Nav-slip-dif-fits}.
For the fitting procedure III it is meaningless
to use GNBC, since the GNBC is equivalent to Navier-slip boundary condition 
for $\mu_f = 0$.
In Fig.~\ref{fig:Nav-slip-dif-fits}(a) we
see the increasing local error of angle near the contact lines, which
remains localised in a thin region near both walls. By investigating
the streamlines (Fig.~\ref{fig:Nav-slip-dif-fits},b-e) we conclude
that indeed as the PF mobility is reduced, the overshoot of the
streamlines in PF is reduced and consequently the agreement with MD
improves. Similar conclusion can be drawn from no-slip PF simulations.

\begin{table}[t]
\begin{center}
\begin{tabular}{p{30mm}|p{20mm}|p{25mm}|p{15mm}}
 & PF no-slip & PF Navier-slip & MD \\ \hline
$\mu_{fa}/\mu_w$ & $0.0$ & $0.0$ & - \\
$\mu_{fr}/\mu_w$ & $0.0$ & $0.0$ & -\\
$\Pe{} = U \epsilon L / \left( M \sigma \right)$ & $2.9$ & $6.0$ & - \\
$\Delta x$ & $5.84$ nm & $5.82$ nm & $5.89$ nm \\
\end{tabular}
\end{center}
\caption{Summary of the obtained phase-field parameters and displacement, fit III.}
\label{tab:maxPe-param-summary}
\end{table}

\section{Volume-of-fluid model} \label{sec:VOF}

The VOF model is known to be well-suited for solving interfacial flows.
The interface between water and vapour in VOF -- in contrast to the PF -- is
reconstructed in a sharp manner.
Therefore the VOF model is a good candidate and is typically used to solve two-phase flows
in macroscopic systems. In this section we investigate how accurately the model
can capture the behaviour of the nanoscopic droplet.

\subsection{Governing equations for the two-phase flow with VOF model}

In the VOF method, the fluid momentum is governed by exactly the same
equations as in the PF method, namely, incompressible Navier--Stokes
equations with variable density and viscosity (\ref{eq:ns-1}--\ref{eq:ns-2}).
The difference in methods lies in three aspects. First, the
governing equation for the concentration function $C$,
which within VOF method is typically called volume fraction, is a
convection equation
\begin{equation}
\frac{\pd C}{\pd t} + \vec{u} \cdot \nabla C = 0,
\label{eq:vof-c-eq}
\end{equation}
instead of the convection--diffusion
equation (\ref{eq:cahn-hil-dim}) for the PF concentration. Second,
the volume fraction $C$ varies from $0$ in the vapour phase to $1$ in the
fluid phase and consequently fluid density and viscosity becomes
\begin{equation}
\rho\left(C\right) = C\,\rho_w + \left(C - 1 \right) \rho_g
\ \ \ \ \mbox{and} \ \ \ \
\mu\left(C\right) = C\,\mu_w + \left(C - 1 \right) \mu_g,
\end{equation}
respectively. The third difference lies in the fact that the surface tension
force within the momentum equation (\ref{eq:ns-1}) takes form
\begin{equation}
\vec{f}_\sigma = \sigma\,\kappa\,\delta_s\,\hat{n},
\end{equation}
where $\kappa$ is the local curvature of the interface, $\delta_{s}$ is
a discrete Dirac distribution function used to spread the surface tension force
to the fluid mesh and $\hat{n}$ is the normal of the two-phase interface.
The curvature and interface normal are approximated using the
Continuum-Surface-Force approach, which states that
\begin{equation}
\kappa \approx \nabla \cdot \hat{n}\ \ \ \ \mbox{and} \ \ \ \
\hat{n} \equiv \frac{\nabla C}{|\nabla C|}.
\end{equation}
The accuracy of the surface tension term is directly dependent on the accuracy
of the curvature calculation. The height-function methodology is a
VOF-based technique for calculating interface normals and curvatures.
The interested reader can find more details on the VOF method 
in appendix~\ref{app:vof}.

\subsection{Boundary conditions for the VOF model}

\begin{figure}[h]
    \centering
    \includegraphics[width=0.45\textwidth]{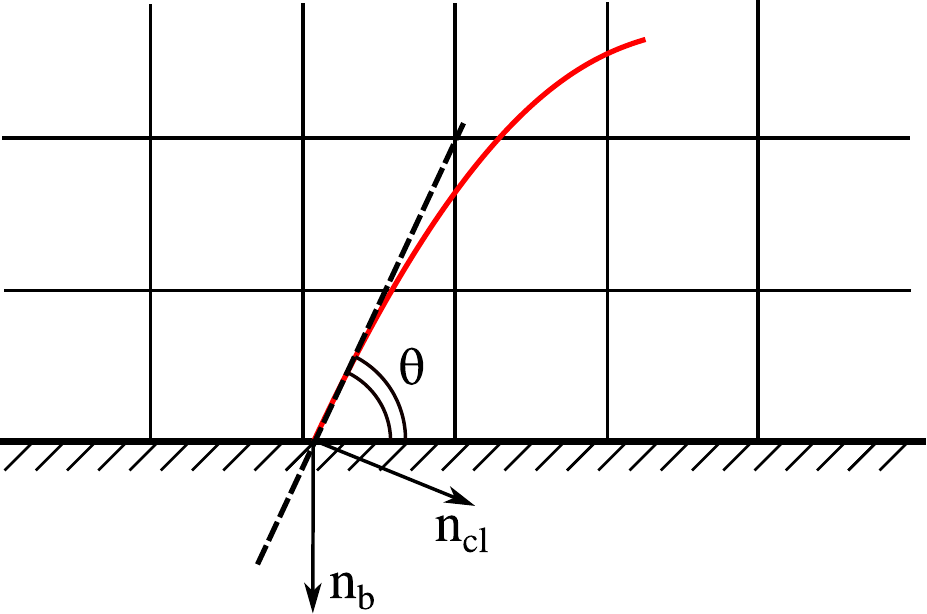}
    \caption{Illustration of wetting condition employed for the VOF model. We use a
    constant contact angle setting that
    defines the normal $\hat{n}_{cl}$ of the interface at the contact line.}
    \label{fig:vof-contactangle}
\end{figure}

For the VOF model used in this work, we use a similar set of boundary conditions
as for the PF model. As we have two sets of unknowns (flow field and volume
fraction), we need boundary conditions to determine both. Due to the fact that we
are interested in steady regime only, we use a constant angle wetting condition.
The angle is imposed through modification of height functions near the boundary
and essentially sets the orientation of interface normal as illustrated in
Fig.~\ref{fig:vof-contactangle}. The imposed contact angle affects
the overall flow
calculation in two ways. First, it defines the orientation of
the VOF interface reconstruction
in cells that contain the contact line and, second, it influences the
calculation of the surface tension term by affecting the curvature
computed in cells at and near the contact line.

The second boundary condition is the velocity condition for the fluid flow.
Although the VOF method does not allow for any diffusive transport of the
contact line, there is always some mesh dependent numerical slip of the
contact line \cite{Afkhami2017}.
This, however, does not provide accurate
control over the results, therefore the explicit specification of a boundary condition
compatible with a moving contact line is desired.
As we have observed from PF result comparison with MD, the Navier-slip condition
alone is sufficient to match the MD results, therefore we restrict the study
on VOF only on the Navier-slip condition. Furthermore, since the observed
slip length from the MD is very small -- recall $l_s = 0.17$ nm -- we made choice
to localise the slip condition in the near vicinity of the moving contact line and
away from the contact line impose the wall velocity $u_x = U_w$.
The localised Navier-slip condition takes form
\begin{equation}
u_x = U_w - f\left( \frac{d}{\epsilon_b} \right)l_s\, \pd_y u_x\, \hat{n}_y. \label{eq:vof-loc-slip}
\end{equation}
Here locality of the slip condition is enforced using the bell function
\begin{equation}
f\left( \frac{d}{\epsilon_b} \right) = \left\{ \begin{array}{cc}
\left[ \frac{1 + \cos \left( \pi d/\epsilon_b \right)}{2} \right]^2 & |d| < \epsilon_b, \\
0 & |d| \geq \epsilon_b.
\end{array} \right.
\end{equation}
The bell function takes as an argument the distance from the contact
line $d = x - x_{CL}$ and the width of the bell function $\epsilon_b$. For coordinates
further than $\epsilon_b$ away from the contact line the function is set to zero
to recover the no-slip condition.
The wall normal velocity component is set to zero $u_y = 0$, same as in PF.
For more details on implementation of the velocity condition in the VOF
model, see appendix~\ref{app:vof}.

\subsection{Comparison with molecular dynamics}

To obtain results from the VOF method, we carry out a fitting
procedure, described as follows.
\begin{enumerate}
  \item Fix the width of the bell function $\epsilon_b = 3.91$ nm or five
  grid sizes in order to capture the variation of the slip length with sufficiently many
  points.
  \item Fix the advancing and receding contact angle to
  $\theta_a = 101^{\circ}$ and $\theta_r = 93^{\circ}$ to represent the
  dynamic angle in the steady regime.
  \item Adjust the magnitude of the local slip length $l_s$ to match
  the drop displacement $\Delta x_m$ from the MD.
\end{enumerate}
Note that for the wetting condition, ideally we would like impose some
relationship or governing law relating contact angle and contact line
velocity. However, because we are looking only at steady regime currently,
we bypass the implementation of a proper
dynamic contact angle model for simplicity.

\begin{figure}[ht!]
\centering
\subfloat[]{\includegraphics[height=5.8cm]{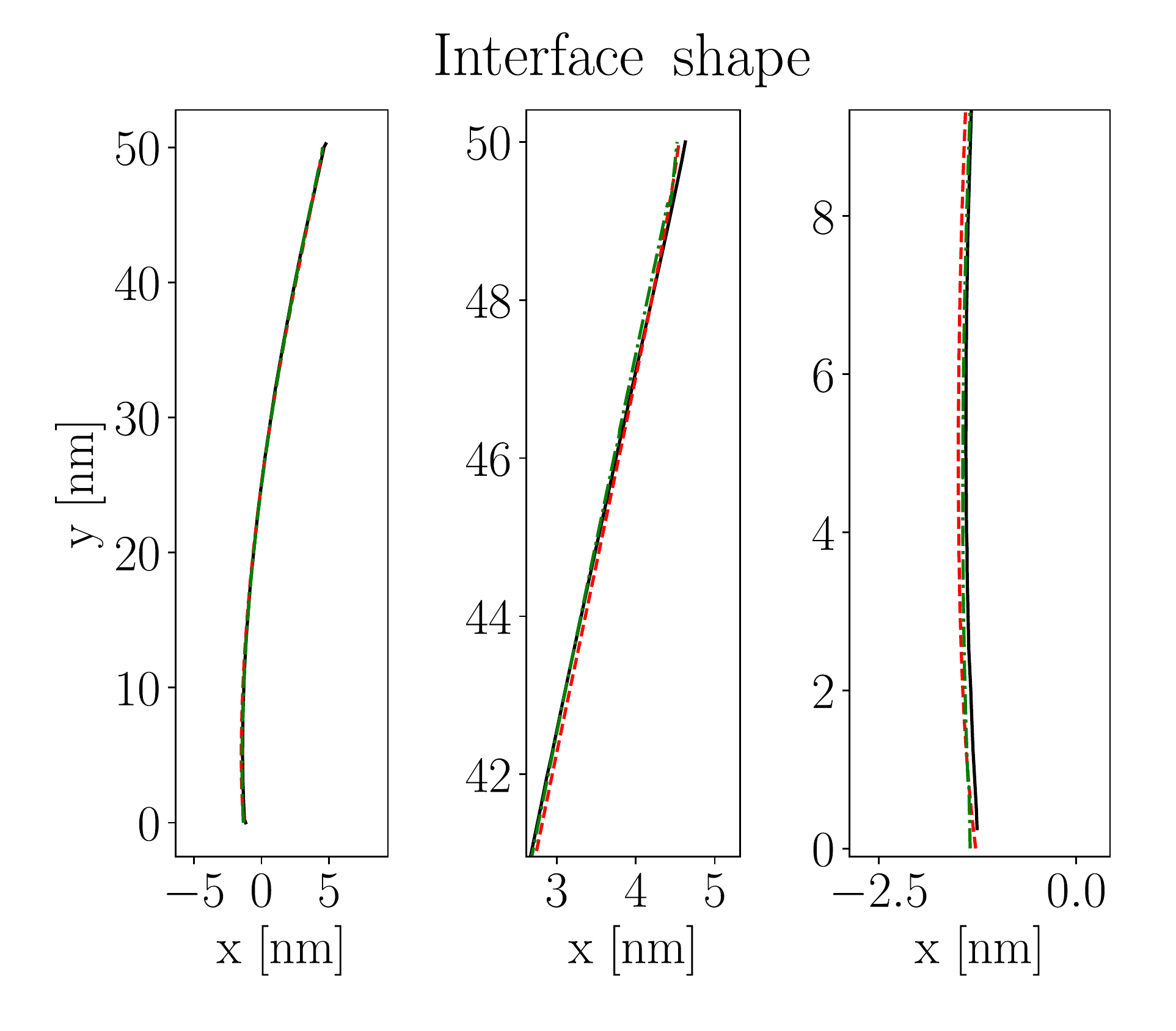}}
\hspace*{10pt}
\subfloat[]{\includegraphics[height=5.8cm]{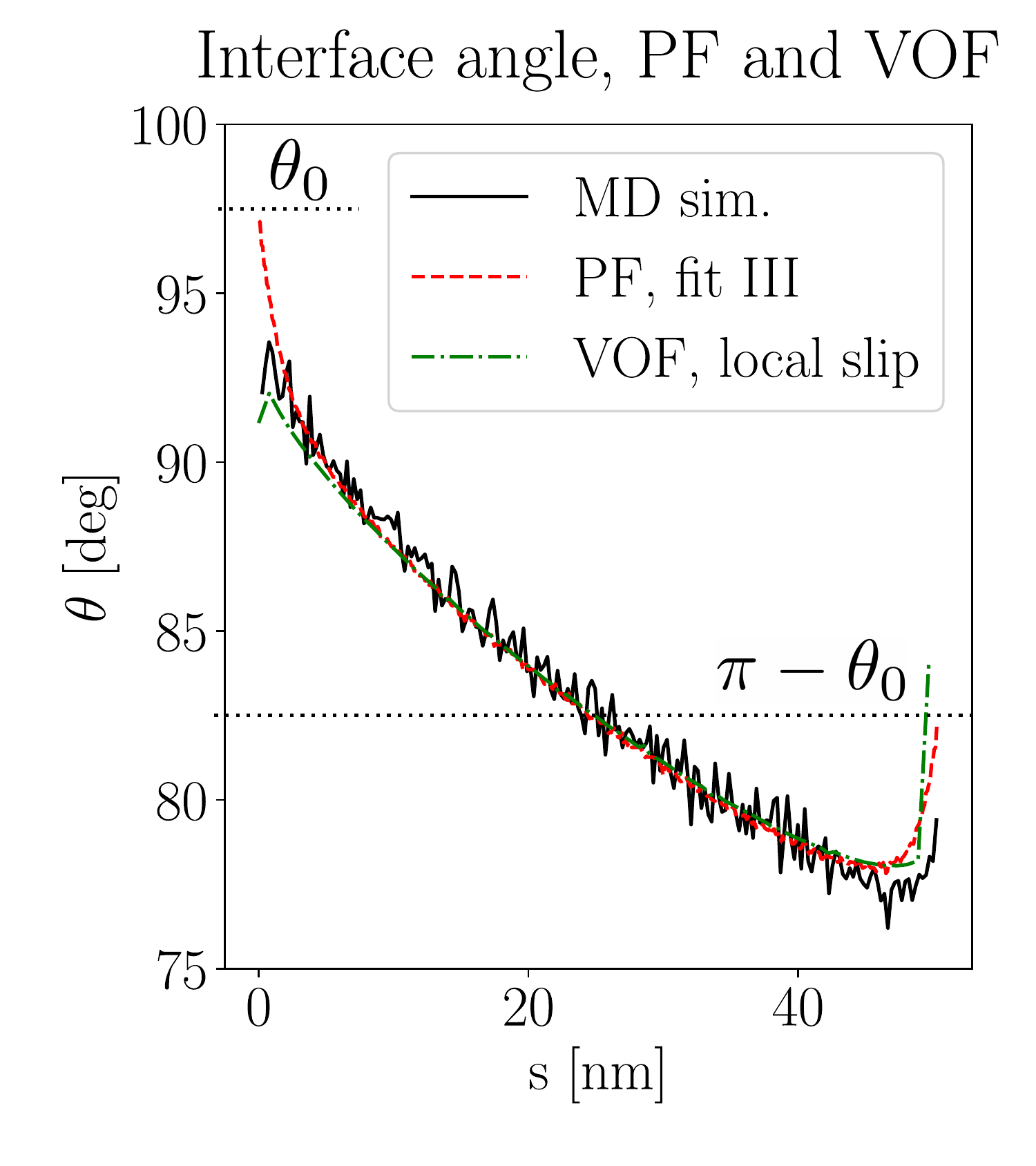}}
\caption{Results of VOF simulations with local Navier-slip boundary condition in
comparison with MD results and best PF results. Interface shape in (a)
and angle distribution over
the interface (b).}
\label{fig:VOF-PF-MD}
\end{figure}

Using this procedure, we have found that the local slip length, which
provides the best match between the VOF and the MD is $l_s = 8$ nm. In
Fig.~\ref{fig:VOF-PF-MD} we show the interface shape and interface angle
comparison between the VOF, PF and MD results. From Fig.~\ref{fig:VOF-PF-MD}(a)
we observe that interface shapes are practically indistinguishable. The interface
angles from VOF, however, show similar local errors as the PF angles. Interestingly,
for the first and last interface angle we observe a rather large jump in the
VOF result, which might be an artefact of the constant contact angle imposition
on the first mesh cell next to the wall.


\section{Discussion} \label{sec:discus}

\subsection{The role of the velocity boundary condition}

As we have observed in the results of the PF simulations, the velocity
boundary condition applied to the fluid momentum equation does not seem to
be important for global measures, such as the drop displacement. Regardless of the
imposed boundary condition, the drop displacement can be captured correctly.
Therefore, for the VOF investigations, we have focused only on local Navier-slip
boundary condition. \rev{For an overview of the fluid slippage conditions
used in different types of simulations
carried out in this work, we provide a summary in Tab.~\ref{tab:allSim-bc-summary}.}

\begin{table}[ht!]
\rev{
\begin{center}
\begin{tabular}{p{30mm}|p{20mm}|p{25mm}|p{23mm}|p{12mm}}
 & PF no-slip & PF Navier-slip & PF GNBC & VOF \\ \hline
slip in vapour phase & $0.0$ nm & $0.17$ nm & $0.17$ nm & $0.0$ nm\\
slip in liquid phase & $0.0$ nm & $0.17$ nm & $0.17$ nm & $0.0$ nm \\
slip at contact line & $0.0$ nm & $0.17$ nm & $0.17$ nm + UYS & $8.0$ nm \\
CLM mechanism & dif. & dif., slip & dif., slip & slip \\
\end{tabular}
\end{center}
\caption{Summary of the simulations carried out in this work, the corresponding
boundary conditions in liquid and gas phases (``UYS'' stands for uncompensated
Young stress), as well as the mechanism of contact
line motion (CLM) -- either diffusion (``dif.''), slippage (``slip'') or a
combination of both.}
\label{tab:allSim-bc-summary}
}
\end{table}

As for the interface shape, which is a more local measure, we have concluded that
both benchmarked continuum models (PF and VOF) had their shortcomings and the
obtained interface
angle always has some local errors near contact line. Notably, the PF has been
successful in describing the interface angle at the receding contact line
(Fig.~\ref{fig:try-perfect-match-itf},b), but it always produced
local error near the
advancing contact line (for $s > 40$ nm, Figs.~\ref{fig:try-perfect-match-itf},b;
\ref{fig:relax-angle-fit},a; \ref{fig:Nav-slip-dif-fits},a;
\ref{fig:VOF-PF-MD},b).
\rev{The possible reasons for this disagreement could be an improperly chosen
wall location as discussed in section~\ref{sec:dis-wall-loc}, and/or a misalignment
of physics between the MD and PF as discussed in section~\ref{sec:dis-diff-PF}.}
Furthermore, when trying to match the
interface angle as close to
the MD as possible, we observe that the
PF model provides non-physical flow field near
the interface (Fig.~\ref{fig:try-perfect-match-flow}), with streamlines crossing
the interface and continuing in the vapour phase. This inaccuracy can be averted
by allowing larger local errors for the interface
angle (Figs.~\ref{fig:relax-angle-fit},a and \ref{fig:Nav-slip-dif-fits},a), which
then reduces the streamline crossing into the vapour phase
significantly (Figs.~\ref{fig:relax-angle-fit},c-e and \ref{fig:Nav-slip-dif-fits},c-e).
Note that the local error can be only observed when plotting the interface angle
(Figs.~\ref{fig:relax-angle-fit},a and \ref{fig:Nav-slip-dif-fits},a), while the
interface shape is globally indistinguishable
(Fig.~\ref{fig:try-perfect-match-itf},a).

The improvement of the flow field was obtained irrespective of chosen boundary
condition (Fig.~\ref{fig:relax-angle-fit},c-e), while the choice of the boundary
condition introduce minor local improvements in either the flow field (if we demand
the same accuracy of local interface angle near contact line) or in the local interface
angle near contact line (if we demand the same accuracy of local flow field, set by
the chosen PF mobility). The reason for improved flow field is that
adding more slippage (first with Navier-slip, and then with GNBC) allows us to
use smaller PF mobility $M$ and consequently obtain streamlines that follow the
two phase interface more accurately.

For the best possible representation of the flow field, we have set the dynamic contact
angle to the equilibrium values, which allowed us to use the smallest PF mobility
parameter. This choice renders the GNBC condition obsolete. By comparing no-slip
and Navier-slip conditions, Navier-slip has produced the most accurate streamlines
with still acceptable representation of the interface shape. However, it is also
possible to use the no slip condition with minor decrease of the flow
field accuracy. Therefore our results indicate that the GNBC condition (or even
Navier-slip condition) is not necessary for physically acceptable
predictions for the flow system considered in the present work using the
PF model. In addition,
increase of the diffusive transport near the contact line seem to always worsen the
flow field prediction, which suggests that sharp interface method could
bet the best possible choice for modelling the present system.

For the accurate representation of the drop displacement in the VOF, we have
fixed the width of the local Navier-slip condition and adjusted the amplitude.
We have observed that using one particular slip length locally near the contact
line allowed us to reach exactly the same drop displacement as MD. However, similar
as for PF, the interface shape has small local
errors in angle (Fig.~\ref{fig:VOF-PF-MD}), while
interface shapes are globally indistinguishable.

\subsection{Wall location and slip condition} \label{sec:dis-wall-loc}

\begin{figure}[ht!]
\centering
\subfloat[]{\includegraphics[width=0.48\linewidth]{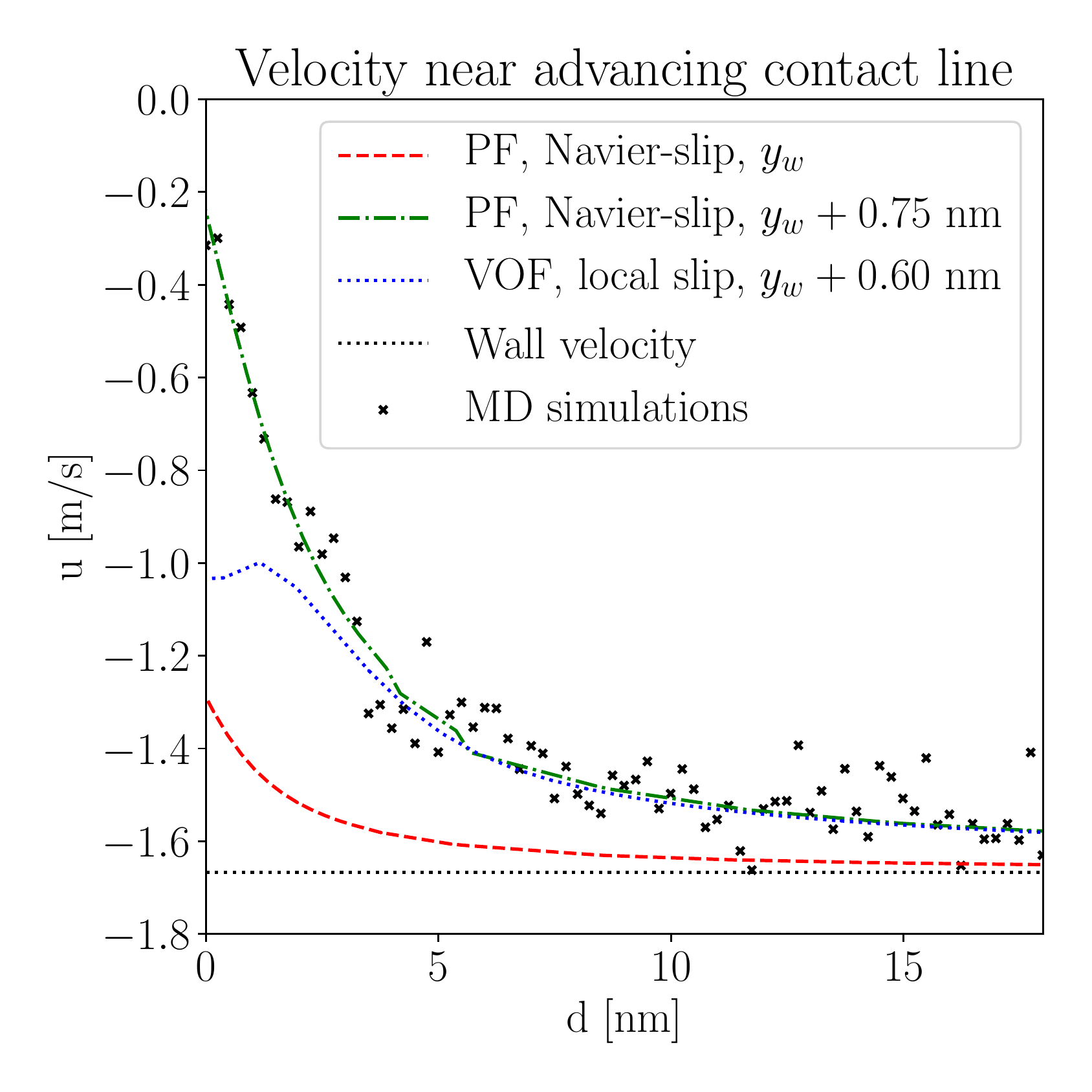}}
\subfloat[]{\includegraphics[width=0.48\linewidth]{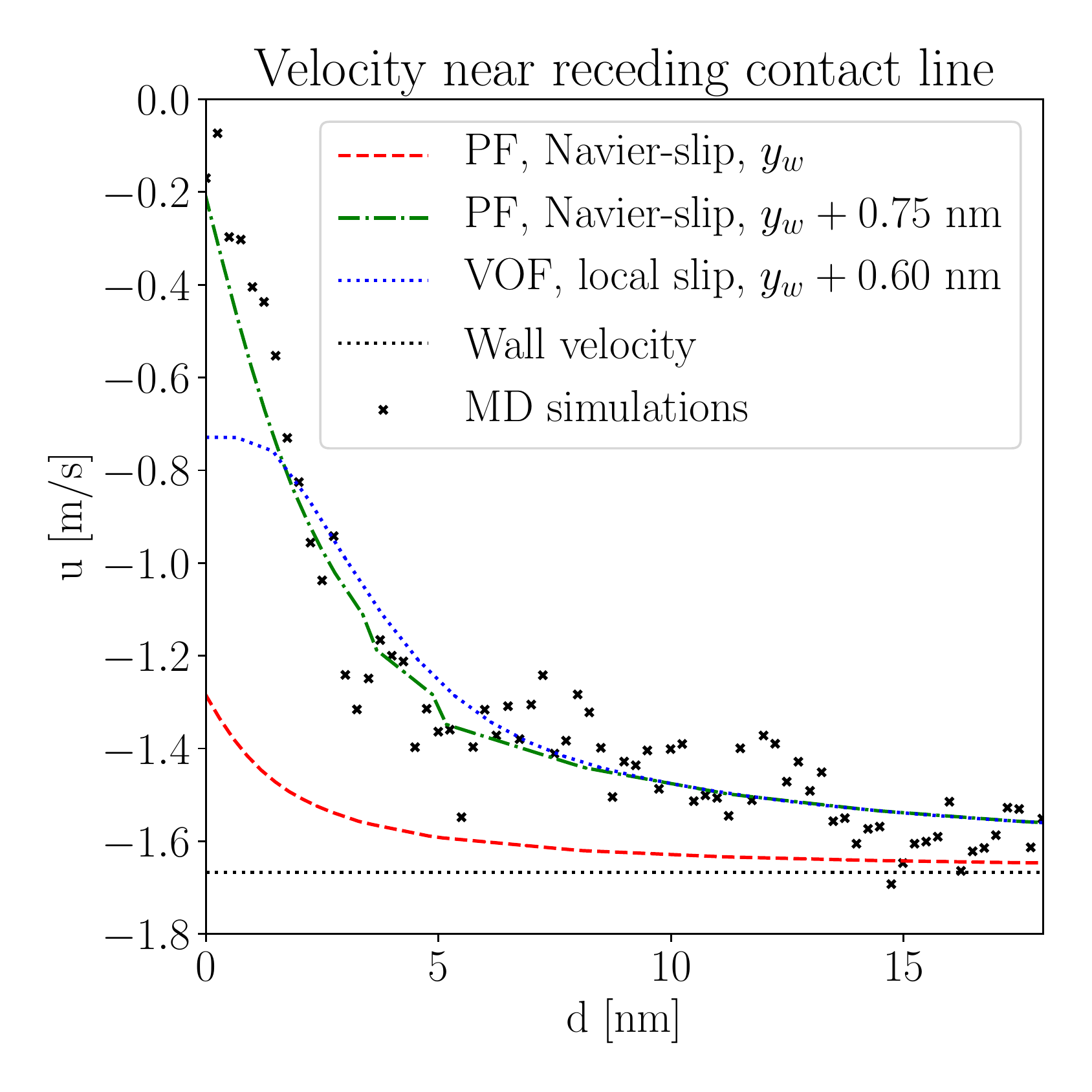}}
\caption{Velocity near advancing (a) and receding (b) contact lines as a function
of $d$ -- the distance from the contact line. MD results are compared with
PF results (providing best streamlines) at two different vertical sampling locations. and
with VOF results at one vertical sampling location.}
\label{fig:vel-near-CL}
\end{figure}

To gain further insight into which boundary condition would be the most appropriate
to model the system consisting of a water drop between two no-slip plates,
we look at velocity distribution as a function of a distance from
the contact line. The MD results are obtained as an average over the 4 runs
centred to the moving contact line.
The MD results are shown in Fig.~\ref{fig:vel-near-CL} with black crosses.

For comparison we show PF with Navier-slip condition,
zero contact line friction $\mu_f = 0$ and $\Pe{} = 6$, which yielded the best fit of 
streamlines. Sampling the velocity field along the wall (Fig.~\ref{fig:vel-near-CL},
red dashed) provides unsatisfactory agreement with the MD both in the near
vicinity of the contact line and also deeper in the drop. However, if the sampling
location is moved $0.75$~nm away from the wall (Fig.~\ref{fig:vel-near-CL},
green dash-dotted), very good agreement is observed
between MD and PF.
Similar observations we have also made
with the VOF method, see Fig.~\ref{fig:vel-near-CL}, blue dotted line, in which
we present VOF results sampled $0.60$~nm away from the wall.
This location provides the best agreement with the PF results away from the wall and
away from the contact line.
Moving the sampling
location up or down would yield a similar effect as demonstrated using the PF. Here, an
interesting observation is that the VOF flow field seems to be less sharp compared
to the PF, which probably stems from the fact that mesh resolution near
the interface in VOF ($\Delta s_{VOF} = 0.782$~nm, appendix~\ref{app:vof-num})
is much coarser than the resolution
of the PF model ($\Delta s_{PF} = 0.195$~nm, appendix~\ref{app:pf2}).

These observations raise two important questions about the presented benchmark. First, the
Navier-slip condition -- based on the slip length measured from the
MD ($l_s = 0.17$~nm) -- does not seem to produce accurate
slip velocity even more than 10~nm away from the contact line. This naturally leads to
a question about where the solid wall in the continuum modelling viewpoint should
be located, compared to the MD molecular picture (Fig.~\ref{fig:md-system-setup}).
The wall location is important for application of the Navier-slip condition,
as displacement of the wall would cause a direct influence on the needed slip length.
\rev{Previous studies \cite{bitsanis1988tractable,hoang2012local} have found that
the liquid structure near a solid surface exhibits layering effect, which is
an additional hint
to a non-trivial correspondence between continuum models and molecular reality.}
The velocity profile agreement at $0.75$~nm distance
from the wall in the continuum simulation seem to suggest that for the best agreement
between MD and PF,
the wall in the PF simulations should be shifted downwards by $0.75$~nm
corresponding to the center of the first bin in Fig.~\ref{fig:md-system-setup}(a).
The other option would be to use much larger slip
length at the wall (larger by a factor of 4 compared to what is currently measured
in the MD). The same discussion
applies also to the VOF results with slightly different
shift of the wall. The slip and wall location therefore
remain open questions.

\subsection{Diffusion of phase-field model} \label{sec:dis-diff-PF}

The P\'{e}clet number in theory exists in the MD system,
because one can define a self-diffusion or a heat diffusion 
coefficient. For the currently used MD water model, the self
diffusivity is \cite{johansson2015water}
\begin{equation}
D_w = 2.3 \cdot 10^{-9} \mathrm{m}^2/\mathrm{s}.
\end{equation}
However, the MD system has a single species and thus
has no mass diffusion. By ``mass diffusion'' we mean that
one species diffuses relative to the other, that is if you consider the mass $m_1$ of one of the species, or its density $\rho_1$, it obeys a diffusion equation.

The PF model in the current work, on the other hand,
is incompressible and consequently does not contain any
state, energy or heat equation and does not model any self
diffusion. The PF model is, however,
based on the diffusion of the phase-field variable $C$,
see equation (\ref{eq:cahn-hil-dim}),
and consequently models mass diffusion of liquid. This
leads to mismatch between the parameters
used in the PF model and diffusion properties determined
from the MD system. The P\'{e}clet number determined
using the self diffusion of the water, relative wall velocity
and height of the water drop is
\begin{equation}
Pe = \frac{2 U_w L}{D_w} = \frac{20 \cdot 50 \cdot 10^{-9}}{6 \cdot 2.3 \cdot 10^{-9}}
= 72.5,
\end{equation}
while maximum value we have been able to use in PF
simulations and represent the MD results accurately
is an order of magnitude smaller, i.e. $Pe = 6$
(see Tab.~\ref{tab:maxPe-param-summary}).

In our opinion, the reason we need to have a moderate $Pe$ number in the PF is twofold; (i) the $Pe$ number can not be tow small because we need to capture the phenomena of very small cross-interface mass transport (equivalent to no evaporation), but on the other hand (ii) $Pe$ number can not be too large because it also controls the amount of diffusive transport very close to the contact line and controls the finally obtained drop displacement. Nevertheless, due to
differences in physical effects MD and PF capture, a perfect
agreement in results should not be expected.

\rev{

\subsection{Density variation of diffuse interface} \label{discuss-den-var}

Related to the PF diffusion, as discussed in
section~\ref{sec:dis-diff-PF}, another interesting question is
the shape of density profile exactly at the interface. In order to gain
an insight in this question, we have performed a smaller MD simulation
(consisting of 10~nm $\times$ 10~nm water droplet)
of exactly the same water model in static conditions, i.e., between two
stationary walls (appendix~\ref{app:molecular-dynamics}).
The obtained density variation at each vertical location has been centred
to half of the water density and the mean has been computed between all the
vertical locations and all data points in time. The result is shown in
Fig.~\ref{fig:density-var-MD-tanh}(a) with
black crosses for $10$~ps averaging window. For this time window the shift 
from vapor to bulk density occurs over $0.7$~nm, matching the  
value of $0.75$~nm reported in section \ref{sec:md-system-parameters} 
where the used bin sizing was $0.25$~nm.

\begin{figure}[ht!]
\centering
\subfloat[]{\includegraphics[width=0.48\linewidth]{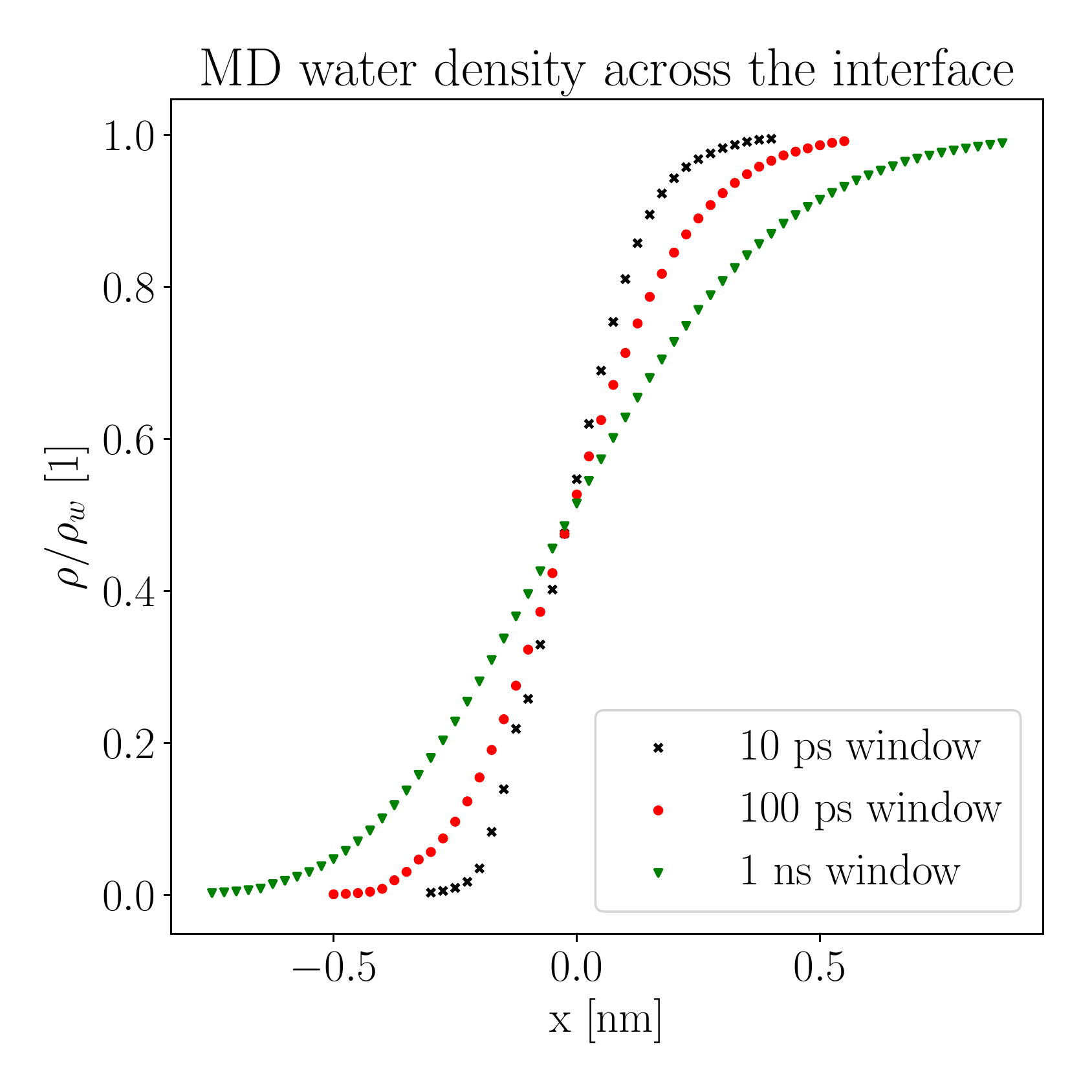}}
\subfloat[]{\includegraphics[width=0.48\linewidth]{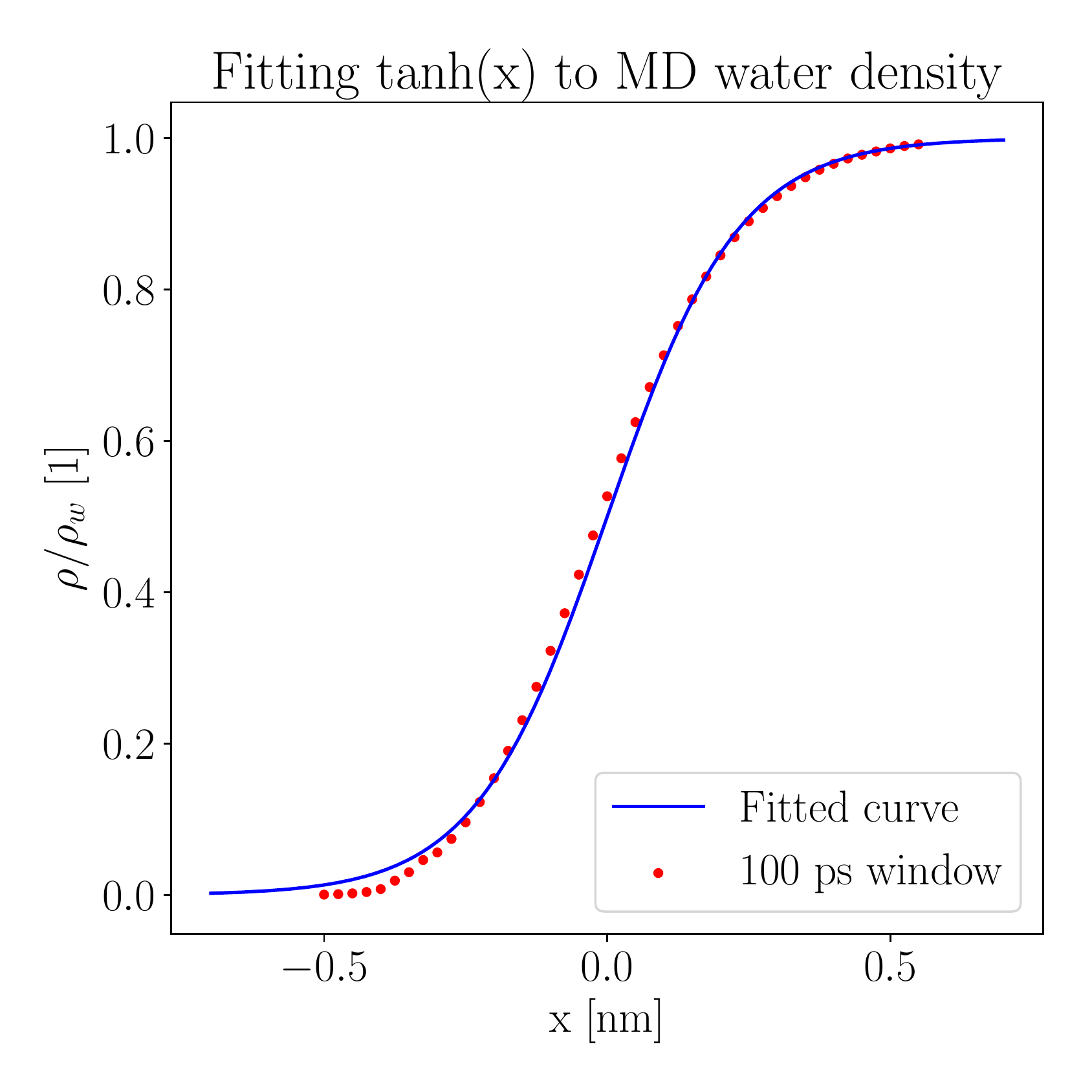}}
\caption{Density variations across the interface from MD with various averaging
windows (a) and PF equilibrium fit to MD data with $100$ ps averaging window. The
MD data is centred at one half of the water density.}
\label{fig:density-var-MD-tanh}
\end{figure}

What this distribution captures for the current MD water model is essentially
thermal fluctuations of the water-vapour interface, as well as small scale
capillary waves. Consequently, this distribution is not invariant with respect
to the averaging window. We illustrate this by showing two more density
distributions in Fig.~\ref{fig:density-var-MD-tanh}(a). When the averaging
window is increased in time, the interface has more time to fluctuate over a
larger distance and consequently the interface appears more diffuse. 

To assess if the diffuse interface in the PF method can be quantitatively
matched with the results we obtained from MD, we recall from the PF
theory~\cite{carlson2012thesis} that in the equilibrium the phase
function $C$ can be expressed analytically as
\begin{equation}
C = \tanh\left( \frac{x}{\sqrt{2} \epsilon} \right),
\end{equation}
where $x$ is distance from the midpoint of the interface. We use this
function in conjunction with the definition of the density (\ref{eq:pf-def-den-visc}),
set the vapour density to zero $\rho_g = 0$ and fit the $\epsilon$ parameter
to match the MD results with $100$ ps averaging window. The obtained fit
is shown in Fig.~\ref{fig:density-var-MD-tanh}(b). The agreement is very good. 

This fit, however,
does not have a direct correspondence to the PF diffuse interface, which in theory
would describe a region where two liquid species are mixed an have an intermediate
density.
In the MD water model, however, the density profile in this interpretation is
perfectly sharp, i.e., there is no region of water/water-vapour mix having intermediate
density. This comparison, however, shows the potential of the PF model
to describe the diffusiveness due to thermal fluctuations and/or capillary waves.
More detailed investigations are needed to determine the extent
of such applicability. For example, it is not clear how large the averaging window
should be chosen and if this interpretations holds all the way down to
the contact line.

}

\section{Conclusions} \label{sec:concl}

We have carried out simulations of steady water drop sheared between two moving plates
using of MD, PF and VOF methods. The MD simulations allow us to probe the
detailed physical picture of the moving contact line and also global measures such
as drop displacement amplitude, interface shape and flow field.

By comparing the results from PF simulations to MD we have observed that a perfect
match is not possible. There is however a choice of parameters and boundary
conditions which provide a reasonable approximation of MD. We observed
a trade-off between accuracy of interface shape and accuracy of the flow field. The best
match with respect to the flow field provided the largest error in the interface shape
and vice versa. Particularly interesting conclusion from this study is that the
exact boundary condition for the flow in the PF simulation does not seem to play a
major role. On the other hand, we have observed that
for accurate predictions there is an upper limit of the Péclet number one can employ ($\Pe{} = 6.0$).
This is an important insight if one considers using PF model as a substitute for
nanoscale simulations, as proposed by Kronbichler and Kreiss \cite{kronbichler2017phase}.
This is also in contrast to previous drop spreading MD and PF
comparisons \cite{johansson2015water}, where $\Pe{} = 1400$ was used. This suggests
that the PF parameters analysed here are not universal and depend on
whether the contact line is advanced through forced wetting or capillary
spreading. We have also concluded, that the GNBC, which in the literature is proposed
as a solution for the moving contact line problem, is not necessary to model the selected configuration.

Based on the results from PF and MD comparison, we have decided to only investigate
a localised Navier-slip boundary condition for the VOF method. With this approach,
we managed to capture the drop displacement accurately, while the interface
shape had similar local errors as observed in the PF model.

By comparing the bottom water layer velocity from MD with the results of PF and VOF, we
have identified an interesting open question about the accuracy of the slip condition
and the wall location. It is possible that from the perspective of the continuum
modelling of such small systems, the wall location might play even more important role
than the chosen boundary condition. We have discussed the
role of diffusion in PF model and the physical phenomena it models, which is
different to what happens in MD simulations,
therefore a perfect agreement should not be expected.
%
%
Another possible future direction would be to investigate
the
width of the localised Navier-slip condition in the VOF model and its influence on
the agreement between VOF and MD. 
Furthermore, comparing the transient behaviour between the MD and continuum
methods would provide additional insight into the robustness of the obtained
continuum parameters.
%

The results of this study will lay rigid foundations for continuum
models of moving contact line
in realistic water/no-slip substrate two phase systems. We believe that this is
the essential first step towards obtaining predictive and robust moving contact
line model, suitable for many real world applications.

\section*{Acknowledgements}

U.L. and S.B. acknowledge funding from Swedish Research Council
(INTERFACE center and grant nr. VR-2014-5680).
S.B. acknowledges funding from 
Knut and Alice Wallenberg Foundation (Grant Number
KAW 2016.0255).
Numerical simulations were performed on resources provided by the Swedish National Infrastructure for Computing (SNIC) at PDC and HPC2N.
U.L. acknowledges Dr. Minh Do-Quang for sharing his femLego code
and assisting with starting up the PF simulations.
All authors acknowledge INTERFACE centre for providing
collaborative environment and making this project possible.
U.L., B.H., G.A. and S.Z. acknowledge the participants of nawet19 workshop for
stimulating and interesting discussions.

\section*{Author contributions}

The idea of this study was conceived by S.B., S.Z., G.A., B.H. and U.L.; the paper was
written by U.L., S.Z., P.J. and T.F. with feedback from all authors. MD simulations
were carried out by P.J., PF simulations were carried out by U.L. and T.F. carried
out VOF simulations. All authors analysed the obtained
results.

\appendix
\renewcommand{\theequation}{\thesection\arabic{equation}}

\section{Details of molecular dynamics simulations}
\setcounter{equation}{0}
\label{app:molecular-dynamics}

The system described in section \ref{sec:molecular-dynamics} and figure \ref{fig:md-system-setup} consists of two parts: the water slab and the moving walls. The water molecules forming the slab are modeled as SPC/E water \cite{berendsen1987}. This is a relatively cheap water model which retains important qualities of real water: the three-point structure and a dipole moment. These features gives the ability for the molecule to form hydrogen bonds with other molecules, which leads to a very low slip length ($l_s = 0.17 \, \textrm{nm}$,
which is smaller than a molecular diameter) and high surface tension.

The walls are built of rigid \SiO{} molecules. Partial charges $q_\textrm{Si} = -2q_\textrm{O}$ are set to the atoms, making the molecules overall charge neutral but with a quadrupole moment. This electrostatic interaction works with the water molecules to create the aforementioned hydrogen bonding. The charge value $q_\textrm{O} = -0.40e$ is used to yield the static contact angle $\theta_0 = \staticContactAngle{}^{\circ}$. The molecules are set in an fcc (111) structure with a spacing of 0.45~nm. They are kept in the desired formation by applying harmonic restraints with spring constant $k = 10\text{,}000 \, \textrm{kJ} \, \textrm{mol}^{-1} \, \textrm{nm}^{-2}$ to the oxygen atoms of each molecule. The molecules are thus free to rotate in the surface normal plane. To move the walls at a desired speed we shift these restraining positions which pulls the molecules along with them while still allowing for natural thermal motion. Remaining details about the wall are described in \cite{johansson2015water}.

Simulations are performed using \textsc{Gromacs 2019} \cite{abraham2015gromacs} in double precision. Atomic positions and velocities are updated using the leap-frog integrator with a time step of 2~fs. Non-bonded van der Waals interactions are treated fully up to a cutoff of 0.9~nm. Coulomb interactions are treated using PME electrostatics which interact over infinite range, including all periodic images of the system. Periodic boundary conditions are applied along $x$ and $z$. Along the $y$ axis reflecting walls are placed at each end to contain molecules to the system, although water molecules are kept from reaching these reflecting walls by the physical \SiO{} walls. A velocity rescaling thermostat is applied to the \SiO{} walls to dissipate excess energy with a time scale of 10~ps and keep the system at simulation temperature $T = 300 \, \textrm{K}$. Outside of the initial 100~ps equilibration period, the thermostat is not applied to the water molecules which can only dissipate energy by interacting with the walls through friction.
The size of the full simulation domain along $x$ is 150~nm, twice that of the water slab.
\rev{We have checked that the dissipation of the heat through walls is sufficient
to keep the average droplet temperature near $T = 300 \, \textrm{K}$.
For evaluating the initial equilibration of the system, we have used the water
vapour pressure and density from the literature \cite{engtoolb_watvapor2004} to
compute the average number of water molecules in the vapour phase at the equilibrium,
which for the current system size yield $15.2$ molecules on average. This is roughly
the amount we see in vapour phase after equilibration. The reason why the system 
equilibrates in such a short time is that the evaporation is a surface
effect.}

Whereas a continuum model can be purely two-dimensional, a molecular system is naturally three-dimensional. A quasi-2D system is created with a system thickness of 4.667~nm in addition to the domain width and height. With this thickness, a width of 75~nm and height 50~nm, the water slab for the simulation is constructed with {$\sim\,$580,000} water molecules.

\rev{
The small simulation used to determine the water density
variation over the interface (section~\ref{discuss-den-var})
is modified as follows.
The binning resolution has been increased from
the previous 0.25~nm $\times$ 0.25~nm (Fig.~\ref{fig:md-system-setup},b)
to 0.025~nm $\times$ 0.025~nm to smoothly capture the density variation.
The data in each bin is sampled over a time interval of $10$~ps and 
the density variations are captured from this data. The simulation is run 
over $32$~ns to obtain sufficient sampling of the density variation, which
is the average of 3200 samples.
To obtain the results for larger average windows, we have combined 10 and
100 of the original samples for $100$~ps and $1$~ns averaging windows, respectively,
in one sample. This construction does not shift the profile of each individual
profile to the center position, and therefore produces more spread of the
interface.
Then final result is obtained from centred average of 320 and 32 samples for
$100$~ps and $1$~ns averaging windows, respectively.

}

\section{Details of phase-field simulations} \label{app:pf}
\setcounter{equation}{0}

In this appendix, we describe in more details the non-dimensional governing
equations as well as numerical implementation of these equations.

\subsection{Dimensionless equations and dimensionless numbers} \label{app:pf1}

In order to render the dimensional governing
equations (\ref{eq:cahn-hil-dim},\ref{eq:ns-1},\ref{eq:ns-2}) and the corresponding
boundary conditions dimensionless, we choose the relative plate velocity $U = 2\,U_w$,
the drop height $L$ and the ration $U/L$ as characteristic scales for velocity,
length and time, respectively. Under this assumption, the convection-diffusion
equation for phase-field variable $C$ (\ref{eq:cahn-hil-dim}) becomes
\begin{equation}
\frac{\pd C}{\pd t} + \vec{u} \cdot \nabla C = \frac{3}{2\sqrt{2}} \frac{1}{\Pe{}} \nabla^2 \phi, \ \
\mbox{with} \ \ \phi = \Psi' \left( C \right) - \Cn{}^2\,\nabla^2 C,
\end{equation}
where all variables are now dimensionless. This procedure gives rise to
P\'{e}clet number and Cahn number,
\begin{equation}
\Pe{} = \frac{U \epsilon L}{M \sigma} \ \ \mbox{and} \ \
\Cn{} = \frac{\epsilon}{L}.
\end{equation}
The P\'{e}clet number is quite important as it provides a measure of relative
importance between convective and diffusive transport of the PF concentration
function $C$. This number we use in the main text to report the needed PF
mobility values for the agreement between PF and MD, see
Tabs.~\ref{tab:fullFit-param-summary}, \ref{tab:relaxAngle-param-summary}
and \ref{tab:maxPe-param-summary}.

The incompressible Navier-Stokes equations (\ref{eq:ns-1}--\ref{eq:ns-2}) in
non-dimensional form becomes
\begin{equation}
\Re{}\,\rho \left( C \right) \left[ \frac{\pd \vec{u}}{\pd t} + \left(\vec{u} \cdot \nabla
 \right) \vec{u} \right] = - \nabla P + \nabla \cdot \left[
 \mu \left( C \right) \left\{ \nabla \vec{u} + \left( \nabla \vec{u} \right)^T \right\}
 \right] - \frac{3}{2\sqrt{2}}\frac{C\, \nabla \phi}{\Cn{}\,\Ca{}},
\end{equation}
where again all variables now are non-dimensional. During this process, two new
dimensionless numbers are introduced, namely Reynolds and capillary numbers, as
\begin{equation}
\Re{} = \frac{\rho_w U L}{\mu_w} \ \ \mbox{and} \ \ \Ca{} = \frac{\mu_w U}{\sigma}.
\end{equation}
The density and viscosity
is normalised with respect to the parameters of water, and therefore in the
dimensionless setting becomes
\begin{equation}
\rho\left(C\right) = \frac{1}{2} \left[ \left( C+1 \right) - \frac{\rho_g}{\rho_w}
\left( C-1 \right)\right], \qquad
\mu\left(C\right) = \frac{1}{2} \left[ \left( C+1 \right) - \frac{\mu_g}{\mu_w}
\left( C-1 \right)\right],
\end{equation}
where $\rho_g$ and $\mu_g$ are the density and the viscosity of the vapour part,
as shown in Fig.~\ref{fig:set-up}.

Finally, the dimensionless wetting condition becomes
\begin{equation}
- \mu_f\,\frac{2\sqrt{2}}{3} \Ca{}\,\Cn{}\, \left( \frac{\pd C}{\pd t} + \vec{u} \cdot \nabla C \right) = \Cn{} \nabla C
 \cdot \hat{n} - \frac{2\sqrt{2}}{3} \cos \left( \theta_0 \right) g'\left( C \right).
\end{equation}
Note that the numerical pre-factors $2\sqrt{2}/3$ appearing in certain places of the
dimensional governing equation stems from relationship between the dimensional
phase-field constants $\alpha$ and $\beta$ and the surface tension,
i.e., $\sigma = 2 \sqrt{2 \alpha \beta} /3$.

\subsection{Numerical implementation} \label{app:pf2}

The dimensionless governing equations introduced in the appendix~\ref{app:pf1} are
linearised, cast into the weak form and solved using a
symbolic finite-element toolbox
\textit{femLego} \cite{amberg1999finite}, which allows easy specification of
finite-element weak form to solve. The solver is based on finite-element package
\emph{deal.II} and includes adaptive mesh refinement to resolve the sharp transition
of phase-field variables over a thin interface \cite{do2007parallel}. Linear
elements were used for the phase-field variables, while the fluid flow was
resolved using Taylor-Hood elements (quadratic for velocity and linear
for pressure).

Mesh resolution was selected after a refinement study based on PF simulation
with a no-slip condition. The final resolution selected and used for all other
simulations is $\Delta s_1 = 3.125$ nm far from the interface, and down to
$\Delta s_2 = 0.195$ nm within the interface region. Constant time step was
used through the simulation as $\Delta t = 0.002$ dimensionless time
units.

\section{Volume-of-fluids model} \label{app:vof}
\setcounter{equation}{0}

In this appendix, we describe in more details the numerical implementation of the solver,
the height functions employed for
the VOF model, as well as the implementation of velocity boundary condition.

\subsection{Numerical implementation} \label{app:vof-num}

In our study of the sheared droplet system, we used the free software \emph{Basilisk}, successor of \emph{Gerris}, developed at the Institut Jean le Rond d'Alembert (Sorbonne Université) \cite{Afkhami2008}, \cite{Popinet2018}, \cite{Afkhami2017}, \cite{Afkhami2009}. The incompressible Navier-Stokes equations are discretised using
the finite volume method and are solved using second order Bell-Colella-Glaz projection
scheme \cite{bell1989second}. The solver is coupled with
VOF method for interface tracking. For obtaining the solution, we have
used a uniform mesh spacing 
of $\Delta s = 0.782$ nm.


\subsection{Height functions}

\begin{figure}[h]
    \centering
    \includegraphics[width=0.45\textwidth]{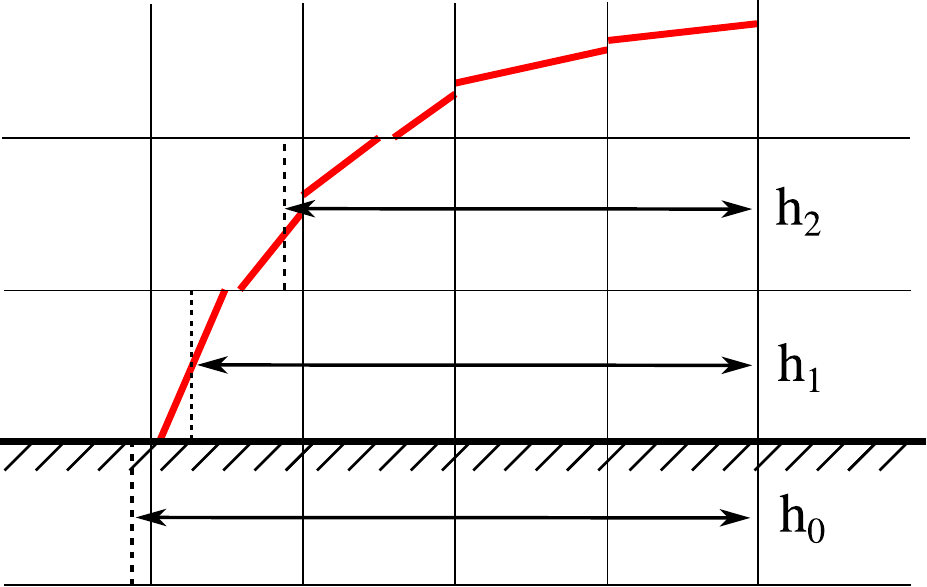}
    \caption{Construction of the 2D height-functions.}
    \label{contactangle}
\end{figure}

The height-function methodology is a VOF-based technique for calculating interface normals and curvatures. About each interface cell, fluid ``heights'' are calculated by summing fluid volume in the direction most normal to the interface. In 2D, a $7 \times 3$ stencil around an interface cell is constructed (Fig.~\ref{contactangle}) and the heights are evaluated by summing volume fractions horizontally, i.e.
\begin{equation}
h_{i}=\sum_{k=j-3}^{k=j+3} C_{i, k} \Delta,
\end{equation}
where $\Delta$ is the grid spacing.
The heights are then used to compute the the interface normal $\hat{n}$ and the
curvature $\kappa$ as
\begin{equation}
\hat{n} = \left(\pd_x h,-1\right) \ \ \mbox{and} \ \
\kappa=\dfrac{\pd^2_{xx} h}{\left(1+(\pd_x h)^{2}\right)^{3 / 2}},
\end{equation}
respectively.
Here, $\pd_{x} h$ and $\pd^2_{xx} h$ are discretised using second-order central differences.

\subsection{Implementation of velocity boundary condition}

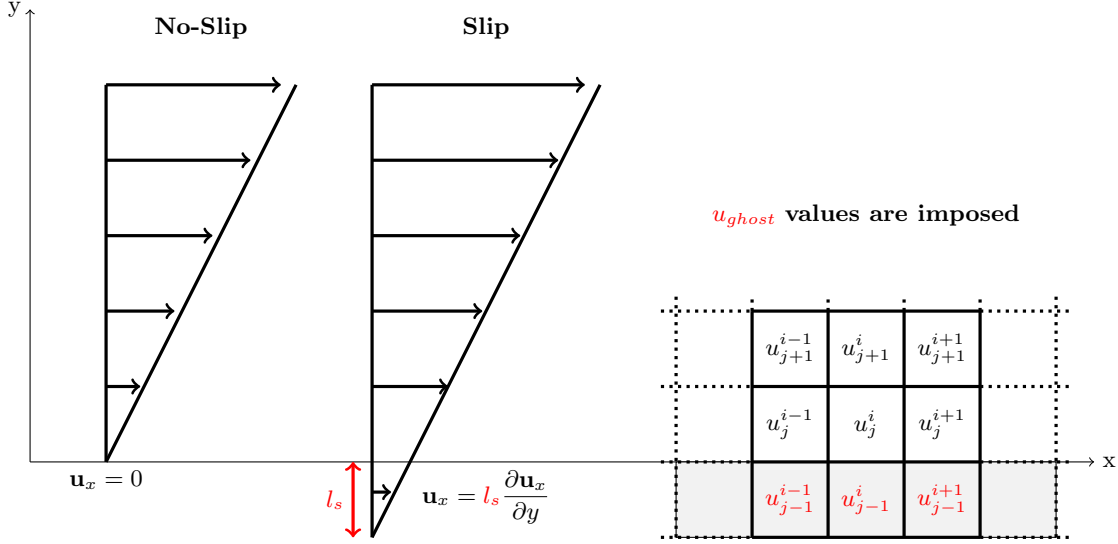
\begin{figure}[h]
\begin{center}
\begin{tikzpicture}
\draw[->] (-5,0) -- (9,0);
\draw (9,0) node[right] {x};
\draw [->] (-5,0) -- (-5,6);
\draw (-5,6) node[left] {y};

\draw[very thick] (-4,0) -- (-4,5);
\draw[very thick] (-4,0) -- (-1.5,5);
\draw[very thick, ->] (-4,1) -- (-3.55,1);
\draw[very thick, ->] (-4,2) -- (-3.1,2);
\draw[very thick, ->] (-4,3) -- (-2.6,3);
\draw[very thick, ->] (-4,4) -- (-2.1,4);
\draw[very thick, ->] (-4,5) -- (-1.7,5);

\draw (-2.75,5.5) node[above] {\textbf{No-Slip}};
\draw (-4,0) node[below] {$\vec{u}_{x} = 0$};

\draw[very thick] (-0.5,-1) -- (-0.5,5);
\draw[very thick] (-0.5,-1) -- (2.5,5);
\draw[very thick, red, <->] (-0.75,-1) -- (-0.75,0);

\draw[very thick, ->] (-0.5,1) -- (0.5,1);
\draw[very thick, ->] (-0.5,2) -- (0.95,2);
\draw[very thick, ->] (-0.5,3) -- (1.45,3);
\draw[very thick, ->] (-0.5,4) -- (1.95,4);
\draw[very thick, ->] (-0.5,5) -- (2.3,5);
\draw[very thick, ->] (-0.5,-0.4) -- (-0.25,-0.4);

\draw[red] (-0.75,-0.5) node[left] {$l_s$};
\draw (1,0) node[below] {$\vec{u}_{x} = \textcolor{red}{l_s} \dfrac{\partial \vec{u}_{x}}{\partial y}$};
\draw (1,5.5) node[above] {\textbf{Slip}};

\draw [fill=gray!10] (3.5,-1) rectangle (8.5,0);

\draw[very thick] (4.5,-1) -- (7.5,-1);
\draw[very thick] (4.5,0) -- (7.5,0);
\draw[very thick] (4.5,1) -- (7.5,1);
\draw[very thick] (4.5,2) -- (7.5,2);
\draw[very thick] (4.5,-1) -- (4.5,2);
\draw[very thick] (5.5,-1) -- (5.5,2);
\draw[very thick] (6.5,-1) -- (6.5,2);
\draw[very thick] (7.5,-1) -- (7.5,2);

\draw[very thick, dotted] (3.3,-1) -- (4.5,-1);
\draw[very thick, dotted] (3.3,0) -- (4.5,0);
\draw[very thick, dotted] (3.3,1) -- (4.5,1);
\draw[very thick, dotted] (3.3,2) -- (4.5,2);

\draw[very thick, dotted] (7.5,-1) -- (8.7,-1);
\draw[very thick, dotted] (7.5,0) -- (8.7,0);
\draw[very thick, dotted] (7.5,1) -- (8.7,1);
\draw[very thick, dotted] (7.5,2) -- (8.7,2);

\draw[very thick, dotted] (3.5,-1) -- (3.5,2.2);
\draw[very thick, dotted] (8.5,-1) -- (8.5,2.2);

\draw[very thick, dotted] (4.5,2) -- (4.5,2.2);
\draw[very thick, dotted] (5.5,2) -- (5.5,2.2);
\draw[very thick, dotted] (6.5,2) -- (6.5,2.2);
\draw[very thick, dotted] (7.5,2) -- (7.5,2.2);

\draw (6, 0.5) node {$u^{i}_{j}$};
\draw (6, 1.5) node {$u^{i}_{j+1}$};
\draw (7, 1.5) node {$u^{i+1}_{j+1}$};
\draw (7, 0.5) node {$u^{i+1}_{j}$};
\draw (5, 1.5) node {$u^{i-1}_{j+1}$};
\draw (5, 0.5) node {$u^{i-1}_{j}$};

\draw[red] (5, -0.5) node {$u^{i-1}_{j-1}$};
\draw[red]  (6, -0.5) node {$u^{i}_{j-1}$};
\draw[red]  (7, -0.5) node {$u^{i+1}_{j-1}$};

\draw (6,3) node[above] {$\textcolor{red}{u_{ghost}}$ \textbf{values are imposed}};

\end{tikzpicture}
\end{center}

\caption[Schematic of velocity profiles for no-slip and slip conditions and the ghost boundary layer. ]{Left side : velocities profiles at the solid interface for the no-slip and slip boundary conditions. Right side : 3x3 stencils with the ghost boundary layer used to impose the boundary condition.}
\label{ghost}
\end{figure}

In order to impose the velocity boundary condition at the wall in the selected
numerical implementation, one has to work with discretisation stencils near the
solid boundary.
These stencils will extend beyond beyond the wall and make use
of so called ghost points (Fig~\ref{ghost}). The stencil values outside the domain (ghost values) need to be initialised. These values are set in order to provide the discrete equivalent of the NBC as
\begin{equation}
\begin{array}{c}
     \dfrac{\vec{u}_{x}[ghost] + \vec{u}_{x}[\:]}{2} + \dfrac{\vec{u}_{x}[ghost] - \vec{u}_{x}[\:]}{\boldsymbol{\Delta}} = U_{w} \\  \\ \iff \vec{u}_{x}[ghost] = \dfrac{2 \boldsymbol{\Delta} }{2 l_s + \boldsymbol{\Delta}} U_{w} + \dfrac{2 l_s - \boldsymbol{\Delta}}{2 l_s + \boldsymbol{\Delta}} \vec{u}_{t}[\:],\end{array}
\end{equation}
where $\vec{u}_{x}[ghost]$ is the tangential velocity at the ghost cell, $\vec{u}_{x}[\:]$ is the tangential velocity of the cell inside the domain and $\boldsymbol{\Delta}$ is the grid spacing.
This implementation is used to impose the locally specified Navier-slip condition,
presented in equation (\ref{eq:vof-loc-slip}).

\bibliographystyle{unsrt}
\bibliography{biblist}


\end{document}

%% file: macros.tex
\newcommand\eps{\epsilon}

\newcommand\be{\begin{equation}}
\newcommand\nd{\end{equation}}
\newcommand\bed{\begin{displaymath}}
\newcommand\ndd{\end{displaymath}}

\newcommand\ba{\begin{array}}
\newcommand\ea{\end{array}}
\newcommand\bea{\begin{eqnarray}}
\newcommand\nda{\end{eqnarray}}

